\documentclass{aastex61}

\newcommand\aastex{AAS\TeX}

\received{}
\revised{}
\accepted{}
\submitjournal{ApJ}

\shorttitle{\aastex\ sample article}
\shortauthors{Zhu et al.}
\usepackage{multirow}

\begin{document}

\title{Determining electron temperature and density in a H II region by using the relative strengths of hydrogen radio recombination lines}

\correspondingauthor{Qing-Feng Zhu}
\email{zhuqf@ustc.edu.cn, zhufya@mail.ustc.edu.cn}

\author[0000-0002-0786-7307]{Feng-Yao Zhu}
\affil{Astronomy Department, University of Science and Technology of China, Hefei, 230026, China}
\affiliation{Key Laboratory for Research in Galaxies and Cosmology, University of Science and Technology of China, Chinese Academy of Sciences, Hefei, 230026, China}

\author{Qing-Feng Zhu}
\affiliation{Astronomy Department, University of Science and Technology of China, Hefei, 230026, China}
\affiliation{Key Laboratory for Research in Galaxies and Cosmology, University of Science and Technology of China, Chinese Academy of Sciences, Hefei, 230026, China}

\author{Jun-Zhi Wang}
\affiliation{Shanghai Astronomical Observatory, Chinese Academy of Sciences, Shanghai, 200030, China}

\author{Jiang-Shui Zhang}
\affiliation{Physics Department, Guangzhou University, Guangzhou, 510006, China}



\begin{abstract}

We have introduced a new method of estimating the electron temperature and density of H II regions by using single dish observations. In this method, multiple hydrogen radio recombination lines of different bands are computed under the assumption of low optical depth. We use evolutionary hydrodynamical models of H II regions to model hydrogen recombination line emission from a variety of H II regions and assess the reliability of the method. According to the simulated results, the error of the estimated temperature is commonly $<13\%$, and that of the estimated density is $<25\%$ for a $<1\%$ uncertainty of the observed line fluxes. A reasonable estimated value of electron density can be achieved if the uncertainty of the line fluxes are lower than $3\%$. In addition, the estimated values are more representative of the properties in the relatively high-density region if the gas density gradient is present in the H II region. Our method can be independent of the radio continuum observations. But the accuracy will be improved if a line-to-continuum ratio at millimeter wavelengths is added to the estimation. Our method provides a way to measure the temperature and density in ionized regions without interferometers.

\end{abstract}




\section{Introduction} \label{sec:intro}

Massive stars form in dense molecular clouds, and affect their parental clouds significantly by feedback processes such as ionizing radiation and stellar wind. After a massive star is born, the extreme ultraviolet ($h\upsilon \geq 13.6$ eV) radiation from the star will ionize the nearby neutral interstellar medium (ISM), and produce an H II region consisting of the ionized gas. Due to the high pressure of the ionized gas, the H II region will expand and compress the surrounding materials enveloping the H II regions. The surrounding compressed materials form a dense shell mainly composed of neutral gases \citep{ten79,mac91}. Besides the ionizing radiation, the stellar wind also has a significant effect on the regions. In an H II region, the stellar wind outflows at a high velocity ($\sim~2000~km/s$) from the center star, and is decelerated by the relatively high-density ionized gas around the star. During the deceleration, the kinetic energy of the stellar wind is converted into thermal energy and a hot stellar and low-density wind bubble ($T~\geq~10^6$K, $n~\leq 1$ cm$^{-3}$) forms inside the photoionized region. Meanwhile, the ionized gas is pushed outward by the over-pressurized bubble, and this increases the expansion velocity of the H II region \citep{com97}.

It is usually assumed that the variations of density and temperature across photoionized region are negligible. If the massive star is born in a uniform environment, the H II region will be spherical and is called a Str$\ddot{\textrm{o}}$mgren sphere \citep{str39}. But, observations show that a large fraction of H II regions are not spherical, but have cometary, pillar-like or other irregular morphologies, and non-uniformity is also obvious and ubiquitous across these H II regions \citep{woo89}. These morphologies could be caused by ambient density distribution or a supersonic stellar motion \citep{ten79,mac91,fra07,mac15}. It has been shown the gas density gradient is important in the formation of these non-spherical H II regions \citep{bod79,ten79b,art06,zhu15b}.

Electron temperature and density are basic properties of H II regions. In a typical H II region, the heating is mainly due to the photoionization, and the cooling is generally dominated by collisionally excited metal lines and hydrogen recombination lines \citep{gae83}. Because the rates of heating and cooling processes have same dependence on the density ($\propto n_e^2$) \citep{spi78}, the equilibrium temperature of the ionized gas is not determined by electron density, but by the chemical abundances and the effective temperature of the center star \citep{pei79,sha83,aff94}. In addition, electron density of H II regions is related to the density of their parental molecular clouds, especially for the young ultra-compact H II regions. So the knowledge of density and temperature in H II regions is helpful to understanding the initial conditions of star formation.

In order to obtain the electron temperatures and densities of H II regions, measurements of intensities of hydrogen recombination lines and free-free radio continuum radiation in different wavebands are always required \citep{woo89,kim17}. Since the optical and infrared observations towards compact H II regions embedded deeply in molecular clouds are hindered due to dust extinction, radio observations are more useful.

In many cases, the electron temperature in H II regions is estimated by measuring the ratio of the hydrogen radio recombination line emission to the continuous radio free-free emission \citep{gor90,gor09}.  The electron temperature estimated by this method is called local thermodynamic equilibrium (LTE) electron temperature $T_e^\ast$ that is usually different from the true value to some extent because of \textbf{departures from} LTE condition. Furthermore, interferometers with a very small beam should be used to obtain the flux density of continuum emission to calculate emission measure (EM) and continuum optical depth if the correct electron temperature and density of a compact H II region are wanted \citep{thu13}. Besides continuum flux density, the flux density of hydrogen recombination lines are also useful for estimating EM and electron density \citep{kim17}. For estimating the electron density, line-of-sight optical depths (LOS depths) of H II regions have to be assumed because the observations usually only provide the information of column density \citep{wil15}. In some cases of ultra-compact H II regions, the effect on the emission line width due to the pressure broadening and the line-to-continuum ratios at multiple wavebands are considered in order to improve the accuracy of estimation \citep{aff94}. For high-density H II regions ($n_e~\geq~10^4~cm^{-3}$), the recombination line widths of high-level ($n>80$) transitions are sensitive to electron density, can provide an accurate estimate of electron density. For low-density H II regions ($n_e~<~3000~cm^{-3}$), since the pressure broadening is small compared to the doppler broadening, the accuracy of this method decreases significantly.

In the previous works mentioned above, the observations of continuum emission are essential for estimating the temperature, and an interferometer generally needs to be used to estimate the electron density. However, the measurement of continuum emission is not available for some telescopes. And the observational time of interferometers is usually limited. In addition, the assumption of LOS depth also introduces uncertainties of the estimated values. These motivate us to find a more feasible way to estimate the properties of the H II regions by single dish observations of multiple hydrogen radio recombination lines.

In the current work, we use the dynamical model \citep{zhu15a,zhu15b} to simulate the evolutions of H II regions with different initial conditions and to calculate several hydrogen radio recombination lines emitted from H II regions in the frequency range between 4.0 and 100 GHz. Through the simulations of these recombination line emissions, the reliability and the limitations of our method are tested. We found the density and temperature of ionized gas can be derived from at least four radio recombination lines. 

The organization of the current work is as follows. In Section \ref{sec:method} we explain our method in details. The results and the discussions about the method used to deal with simulated H II regions are showed in Section \ref{sec:result}. In Section \ref{sec:conclusion} the conclusions are presented.


\section{Numerical Method} \label{sec:method}

\subsection{line intensity ratios \label{sec:intensity}}

The emission coefficient of a hydrogen recombination line per unit volume is as follow \citep{gor09}:

\begin{equation}
j_{\nu,L}=\frac{h\nu}{4\pi}\phi_\nu N_mA_{m,n}~~~~,
\end{equation}

where $N_m$ is the number density of hydrogen atoms with the principal quantum number $m$, $A_{m,n}$ is the spontaneous Einstein coefficient with upper state $m$ and lower state $n$, and $h$ is the Plank constant. $\phi_\nu$ is the normalized line profile \citep{pet12}. In this paper, we only discuss lines in the $\alpha$ series. 
On the assumption of local thermodynamic equilibrium, the line emissivity can also be calculated as

\begin{equation}
j_{\nu,L}=B_{\nu}(T)\kappa_{\nu,L}\approx\frac{2kT\nu^2}{c^2}\kappa_{\nu,L}~~~,~h\nu\ll kT,
\end{equation}

where $B_{\nu}(T)$ is the intensity of a blackbody at the temperature $T$ and frequency $\nu$, and $\kappa_{\nu,L}$ is the line absorption coefficient. $c$ the speed of light. The free-free continuum emissivity is

\begin{equation}
j_C=B_{\nu}(T)\kappa_{\nu,C}
\end{equation}

where $\kappa_{\nu,C}$ is the continuum absorption coefficient \citep{ost61,dic03}. The total intensity of the line and continuum emissions on LTE assumption provided in \citet{gor09} is

\begin{equation}
I_\nu^{LTE}=B_{\nu}(T)[1-e^{-(\tau_{\nu,C}+\tau_{\nu,L})}].
\end{equation}

And if the continuum and line optical depths $\tau_{\nu,C}$ and $\tau_{\nu,L}\ll1$, the intensity of the line emission in LTE is

\begin{equation}
I_{\nu,L}^{LTE}\approx B_\nu(T)\tau_{\nu,L}=\int \frac{h\nu}{4\pi}\phi_\nu N_m^{LTE}A_{m,n}dD~~~~, h\nu\ll kT_e,
\end{equation}

with

\begin{equation}
N_m^{LTE}=\frac{n_en_i}{T^{3/2}}\frac{m^2h^3}{(2\pi m_ek)^{3/2}}exp(\frac{E_m}{kT})~~~,
\end{equation}

where $D$ is the line-of-sight depth. $T_e$ is electron temperature. $n_e$ and $n_i$ are electron and ion number densities, respectively. In non-LTE situation, and also under the assumption of low optical depth, the intensity of the line emission needs to be corrected as

\begin{equation}
I_{\nu,L}\approx I_{\nu,L}^{LTE}b_m(1-\frac{\tau_{\nu,C}}{2}\beta)~~~,
\end{equation}

where $b_m$ is the departure coefficient in energetic level $m$ determined by $T_e$ and $n_e$. $\beta$ is a function of $T_e$ and departure coefficients as below \citep{gor09}

\begin{equation}
\beta=\frac{1-(b_m/b_n)e^{-h\nu/kT_e}}{1-e^{-h\nu/kT_e}}~~~.
\end{equation}

Then the frequency-integrated intensity is

\begin{equation}
\int I_{\nu,L}d\nu\approx \int b_m(1-\frac{\tau_{\nu,C}}{2}\beta)\frac{h\nu}{4\pi}\phi_\nu N_m^{LTE}A_{m,n} dDd\nu~~~.
\label{eq_rbeta}
\end{equation}

The continuum optical depth $\tau_{\nu,C}$ is a function of electron temperature $T_e$, electron number density $n_e$ and the LOS depth $D$. Then a ratio of frequency-integrated intensities of two hydrogen recombination line is determined by $T_e$, $n_e$ and $D$. To estimate three unknown parameters, at least four hydrogen recombination lines (three ratios) are needed. In addition, the parameter $\beta$ is negative, and its absolute value is usually in the range of $1$ to $100$ for H II region gas at millimeter and centimeter wavelengths. If the continuum optical depth is even low as $\tau_{\nu,C}<0.001$, the frequency-integrated line emission intensity can be written as

\begin{equation}
\int I_{\nu,L}d\nu\approx \int b_m\frac{h\nu}{4\pi}\phi_\nu N_m^{LTE}A_{m,n}dDd\nu~~~.
\label{eq_rI}
\end{equation}

In this case, the ratio of intensities of two hydrogen recombination lines is only determined by $T_e$ and $n_e$. Only three hydrogen recombination lines (two ratios) are essential for estimation. Derived from Eq. \ref{eq_rI}, the ratio of two departure coefficients with principal quantum number $m_1$ and $m_2$ can be presented as

\begin{equation}
\frac{b_{m_1}}{b_{m_2}}\approx\frac{\int I_{\nu_1,L}d\nu}{\int I_{\nu_2,L}d\nu}\frac{\nu_2m_2^2A_{m_2,n_2}}{\nu_1m_1^2A_{m_1,n_1}}~~~,E_{m1}~\textrm{and}~E_{m2} \ll kT,
\end{equation}

The population and departure coefficients of hydrogen atoms on excited levels in ionized gas has been studied for decades \citep{sej69,gor90}. They could be calculated by the so called n-model considering bound-bound and bound-free radiative transitions as well as inelastic collisions \citep{sej69,bro70,bur76,wal90}. The more sophisticated nl-model including angular momentum changing collisions was used in \citet{sto95}. Recently, \citet{pro18} presented new results using nl-model with new numerical techniques to ensure convergence of the solution and updated calculations about radiative and collisional transitions, and the effects on departure coefficients of continuum radiation are also considered.

\begin{figure}[ht!]
\centering
\includegraphics[scale=0.45]{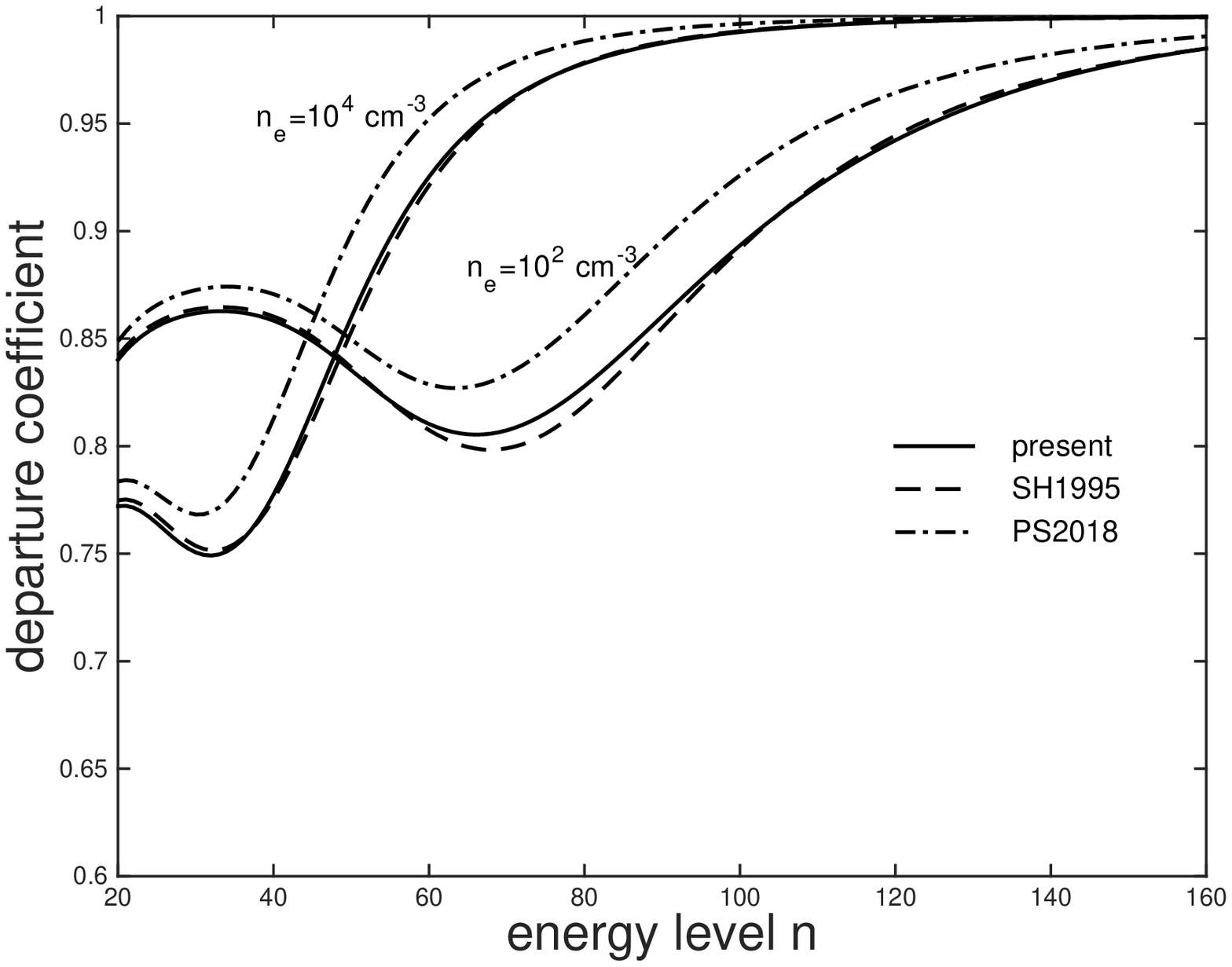}
\includegraphics[scale=0.45]{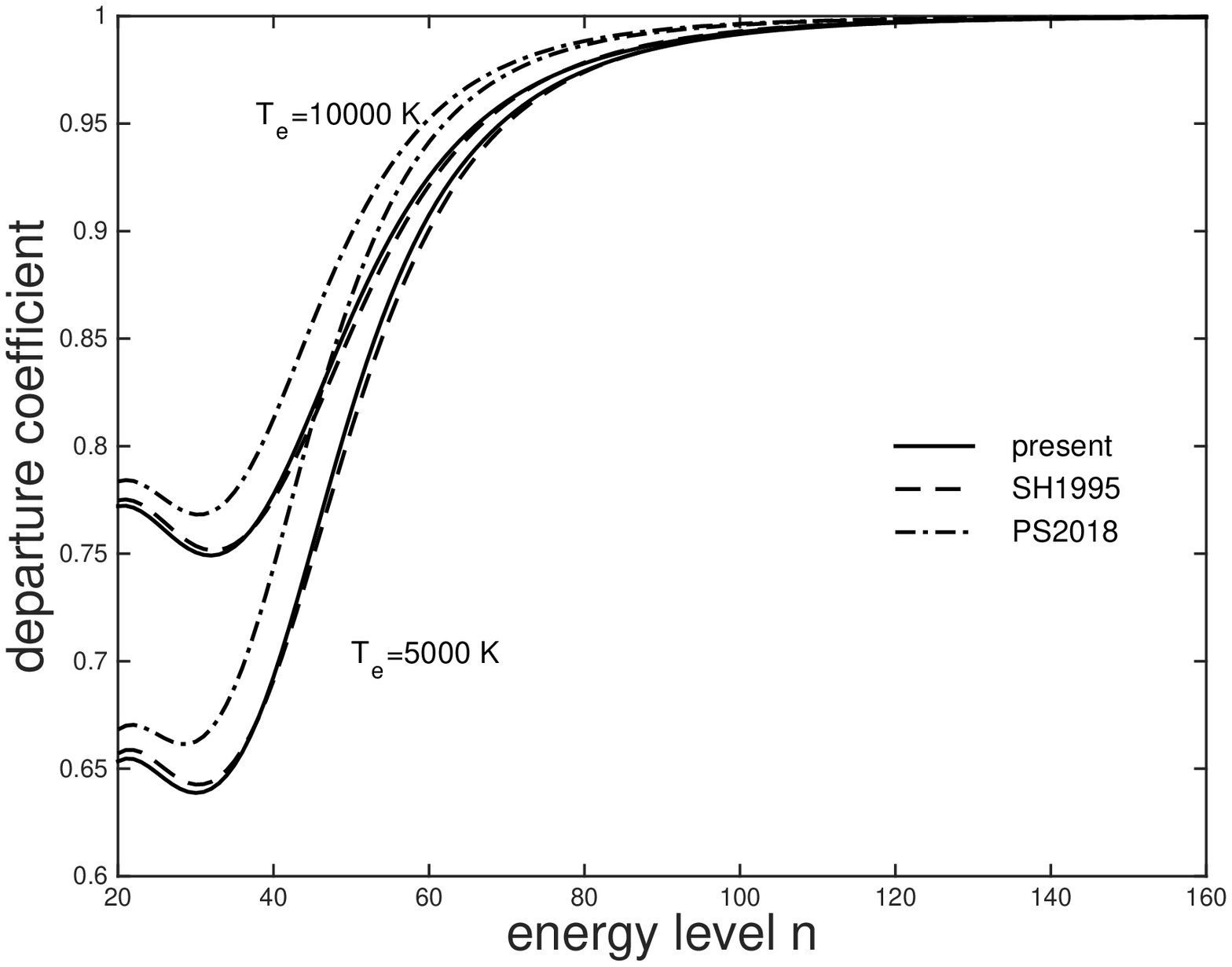}
\caption{the variations of the departure coefficients with $T=10000~K$ and $n_e=100$ and $10000~cm^{-3}$ are presented in the left panel, and those with $n_e=10000~cm^{-3}$ and $T=5000$ and $10000~K$ are presented in the right panel. The results provided by \citet{sto95} and  \citet{pro18} are marked as 'SH1995' and 'PS2018', respectively. These results are all calculated without external radiation fields.}
\label{fig:departure}
\end{figure}

In order to obtain correct hydrogen recombination line strengths, the accuracy of the calculation about the departure coefficients $b_m$ for the occupation number density is important. We used an algorithm similar to the nl-method given by \citet{pro18} to compute the values of $b_m$. When calculating the departure coefficients, several important processes affecting the level populations including radiative recombination and cascade, collisional excitations and deexcitations induced by electrons and protons, collisional ionization, three-body recombination, and angular momentum changing collisions are considered as in \citet{pro18}. But the assumption of case B approximation in our method is treated as a 'standard Case B' described in \citet{hum87}.  \textbf{All Lyman transitions are assumed to be optically thick so that the Lyman photons are absorbed on-the-spot. In addition, the collisional excitations from n=1 and n=2 states are ignored.} The stopping criterion suggested in \citet{pro18} is not used, and we terminate the iterative procedure if the largest difference of $b_m$ between two iterations is less than $1.0\times10^{-6}$. \textbf{The radiation field could also affect the level populations, but this effect become noticeable only if  $n_e>10^4~cm^{-3}$ \citep{pro18}. This condition is not satisfied by the H II region models simulated in the current work.} So the effect of radiation field is neglected in our algorithm. In order to assess the reliability of the algorithm to calculate departure coefficients, the results given by \citet{sto95} and \citet{pro18} are compared with ours. The comparison is presented in Figure \ref{fig:departure}. In the range of principal quantum number n=20 to 160 and at the condition of T=10000 K and n$_e$=10000 cm$^{-3}$, the average deviation between our values and those of \citet{sto95} is $0.29\%$. And compared with those of \citet{pro18}, it is $2.0\%$. The largest discrepancies are $0.79\%$ and $4.7\%$ compared with \citet{sto95} and \citet{pro18}, respectively. It is surprising that our results are closer to those of \citet{sto95} than to those of \citet{pro18} although we followed the calculation method provided by the latter. The different values can not be fully explained only by the differences in case B approximation and the stopping criterion. The real reason is still unknown.

\begin{figure}[ht!]
\centering
\includegraphics[scale=0.45]{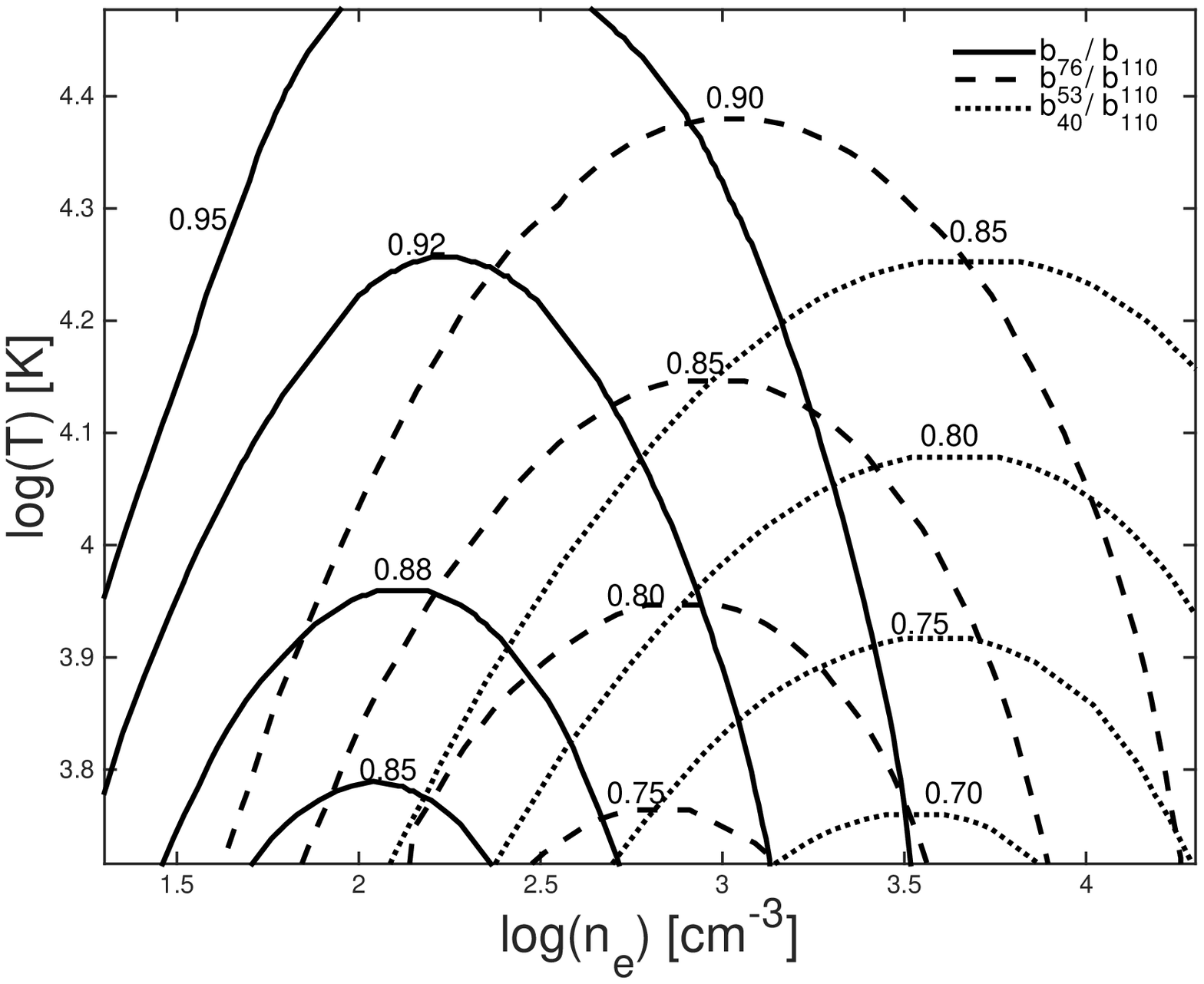}
\includegraphics[scale=0.45]{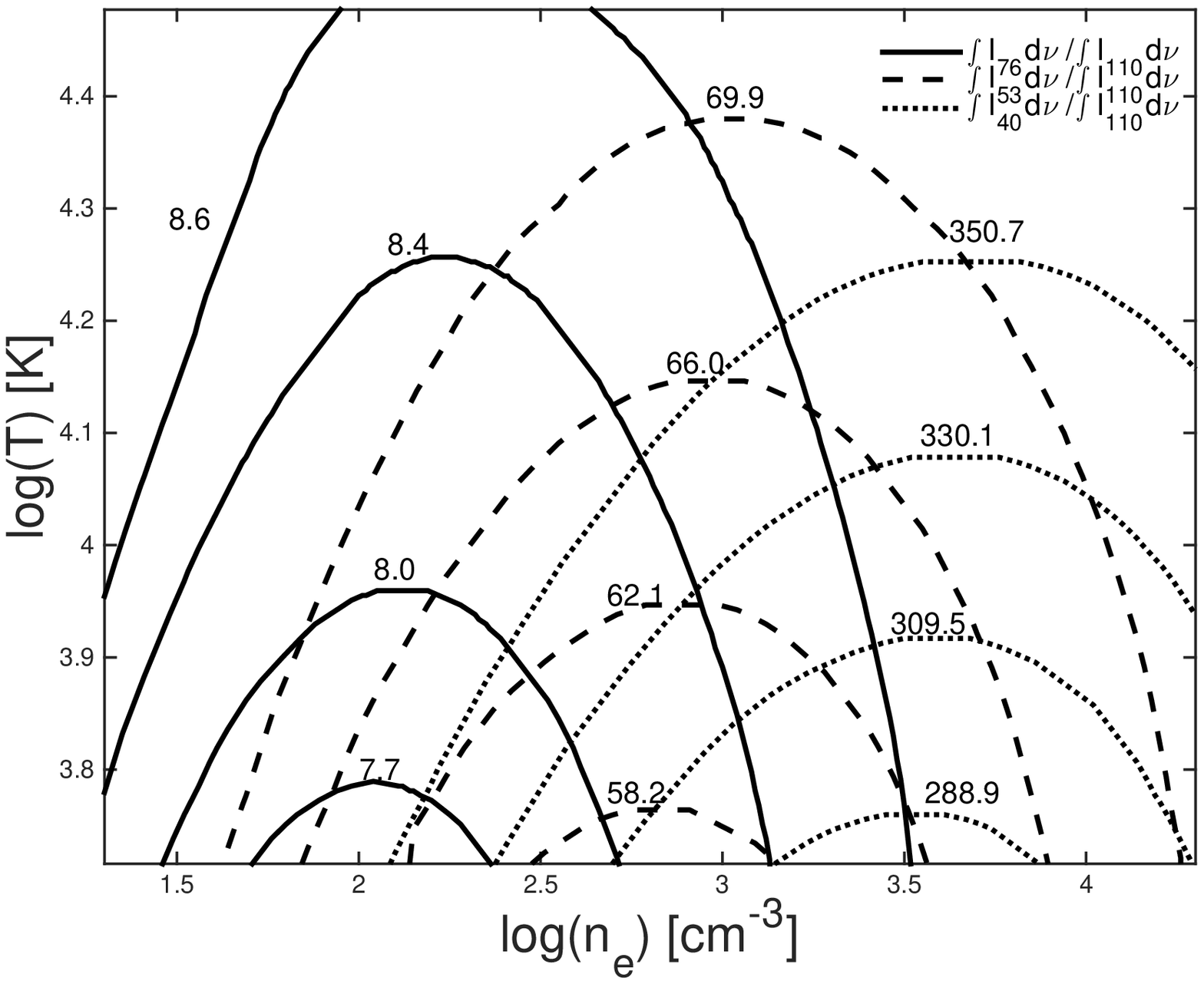}
\caption{Lines of constant departure coefficient ratios are shown for four radio bands in the left panel. The temperature and density of the ionized gas can be estimated from the crossing of the lines. The corresponding lines of constant intensity ratios are presented in the right panel, and $\int I_{76}d\nu$ means the frequency-integrated intensity of H76$\alpha$ line. Hn$\alpha$ lines are labelled by the principal quantum number n at the lower state of the transition.}
\label{fig:dratios}
\end{figure}

\begin{figure}[ht!]
\centering
\includegraphics[scale=0.6]{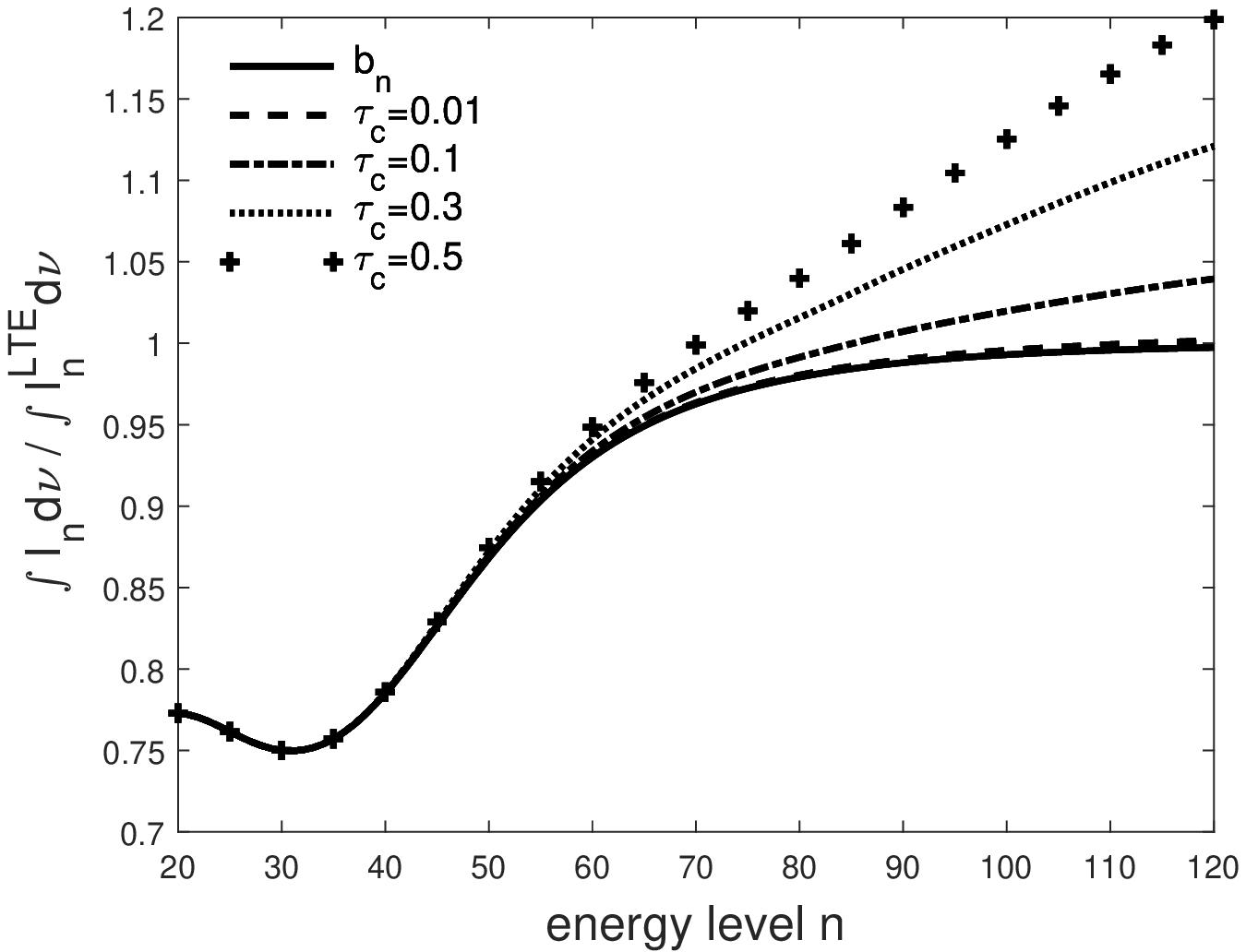}
\includegraphics[scale=0.6]{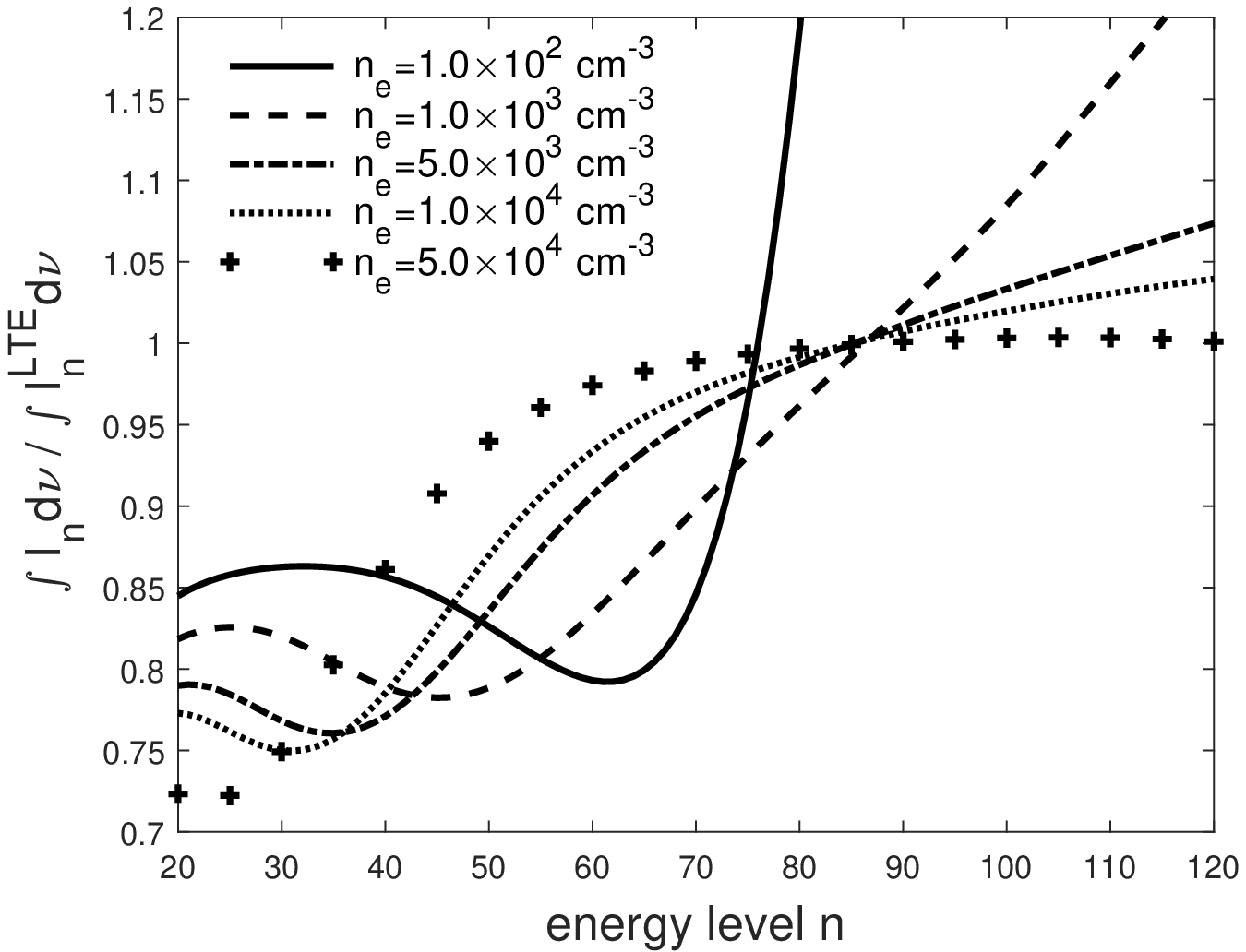}
\caption{The variationis of the ratios of the frequency-integrated intensities to the LTE frequency-integrated intensities of Hn$\alpha$ lines. The curves for $T_e=10,000~K$ and $n_e=10,000~cm^{-3}$ with different continuum optical depths at 5 GHz ($\tau_C$) are presented in the left panel. The curves for $T_e=10,000~K$ and $\tau_C=0.1$ at 5 GHz (EM$\approx1.0\times10^7~cm^{-6}pc$) with different $n_e$ are plotted in the right panel. $b_n$ means the values of departure coefficients. $\int I_nd\nu$ means the frequency-integrated intensity of an Hn$\alpha$ line, and $\int I_n^{LTE}d\nu$ is the LTE intensity.}
\label{fig:bbeta}
\end{figure}

In the left panel of Figure \ref{fig:dratios}, we present isolines of departure coefficient ratios ($b_{41}/b_{111}$, $b_{54}/b_{111}$ and $b_{77}/b_{111}$) in the temperature-density diagram. $T_e$ and $n_e$ of the ionized gas emitting hydrogen recombination lines correspond to the position where two isolines cross. In the right panel, the corresponding isolines of intensity ratios are plotted by using Eq. \ref{eq_rI} to calculate the relative strengths of lines under the assumption that $\tau_{\nu,C}<0.001$. Because of the small difference between the departure coefficients associated with neighbouring levels, it is necessary to observe hydrogen recombination lines in different bands. Better estimates can be obtained by including more hydrogen recombination lines. The least square method can be used when more than 4 hydrogen recombination lines are added to estimation. $T_e$ and $n_e$ will be underestimated if the assumption of very low $\tau_{\nu,C}$ is not satisfied. In this case, $T_e$ and $n_e$ can be estimated by replacing the departure coefficients with the ratio of $\int I_{v,L}d\nu$ to $\int I_{v,L}^{LTE}d\nu$ in fitting. The ratios of $\int I_{v,L}d\nu$ to $\int I_{v,L}^{LTE}d\nu$ of Hn$\alpha$ lines at $T_e$ of $10,000~K$ and $n_e$ of $10,000~cm^{-3}$ with different continuum optical depths at 5 GHz are presented in the left panel of Figure \ref{fig:bbeta}. \textbf{The intensities are calculated by using Eq. 22 and 25 in \citet{pet12} without optically thin assumption since the continuum optical depths could be higher than 0.1.} It is showed that the intensities of the hydrogen recombination lines increases obviously at centimeter wavelengths with the rising continuum optical depth. But the increase is very small at millimeter wavelengths. The ratios for $\tau_{\nu,C}=0.1$ with $\nu=5.0$ GHz (EM$\approx1.0\times10^7~cm^{-6}pc$) and $T_e=10,000~K$ with different $n_e$ values are plotted in the right panel. It is presented that the increase of the ratio is more obvious in the condition with low electron density. From Figure \ref{fig:departure} and \ref{fig:bbeta}, the position of the trough in the curves about $b_n$ and $\int I_{v,L}d\nu/\int I_{v,L}^{LTE}d\nu$ seems to be an indicator of $n_e$, and it is not sensitive to $T_e$.



\subsection{the hydrodynamical model test \label{sec:departure}}

We can test the above method by using the hydrodynamical and radiative transfer model of H II region. In the dynamical model, the hydrodynamics and the radiative transfer equations are calculated at the same time, and the gravity and magnetic fields are neglected. The evolution of the H II region and its surrounding PDR is treated with a 2D explicit Eulerian method in cylindrical coordinates. The simulations of H II regions are all computed on a grid of 250 radial by 500 axial direction cells. The cell size of the grid is set according to the size of the corresponding simulated H II region. The other details of the H II region model is provided in \citet{zhu15a} and \citet{zhu15b}. After physical and chemical quantity distributions of H II regions are simulated by the dynamical model, the properties of the hydrogen recombination lines are calculated. Then the calculated characteristics of these lines are used to estimate the electron temperature and density of H II regions.

\section{results and discussions} \label{sec:result}

In order to test the feasibility of using multiple hydrogen radio recombination lines to estimate the electron temperature and density of an H II region, we simulate several series of models of H II regions with different initial conditions and evaluate the results in these models. The initial conditions of the models are listed in Table \ref{table_name}.

\begin{table}[h]
\centering
\begin{tabular}{|c|c|c|c|c|}
\hline
Name & Morphology & $M_\ast$ [$M_\odot$] & $n_H$ [$cm^{-3}$] & $v_\ast$ [$km~s^{-1}$] \\
\hline
A1 \& B1 & spherical & 19.0 & 3000 & 0 \\
A2 \& B2 & spherical & 19.0 & 5000 & 0 \\
A3 \& B2 & spherical & 19.0 & 10000 & 0 \\
A4 \& B4 & spherical & 19.0 & 15000 & 0 \\
A5 \& B5 & spherical & 19.0 & 20000 & 0 \\
\hline
C1 & spherical & 40.9 & 3000 & 0 \\
C2 & spherical & 40.9 & 5000 & 0 \\
C3 & spherical & 40.9 & 10000 & 0 \\
C4 & spherical & 40.9 & 15000 & 0 \\
C5 & spherical & 40.9 & 20000 & 0 \\
\hline
D & non-spherical & 40.9 & 8000 & 15 \\
E & non-spherical & 40.9 & $5.0\times10^4exp(z/0.05pc)$ & 0 \\
F & non-spherical & 40.9 & $4.0\times10^6(r/0.05pc)^{-2}$ & 0 \\
\hline
\end{tabular}
\caption{The initial conditions of the spherical H II region models. Compared to Model A series, a stellar wind is considered in Model B series. The stellar masses are increased with stronger stellar wind and ionizing radiation in Model C series. $M_\ast$ is stellar mass, and $v_\ast$ means the stellar velocity. $n_H$ is the initial number density of uniform interstellar medium, and the variation of the density against coordinate is displayed if there is a density gradient in non-spherical models.}\label{table_name}
\end{table}

\subsection{properties of a spherical H II region with $EM<1.0\times10^5~cm^{-6}pc$} \label{sec:set}

In the first, we apply the method to the H II regions with very low continuum optical depth of $\tau_{\nu,C}<0.001$ at 5 GHz ($EM<1.0\times10^5~cm^{-6}pc$) so that the parameter $b_m(1-\tau_{\nu,C}\beta/2)$ is approximately equal to the departure coefficient $b_m$.

\begin{figure}[ht!]
\centering
\includegraphics[scale=0.5]{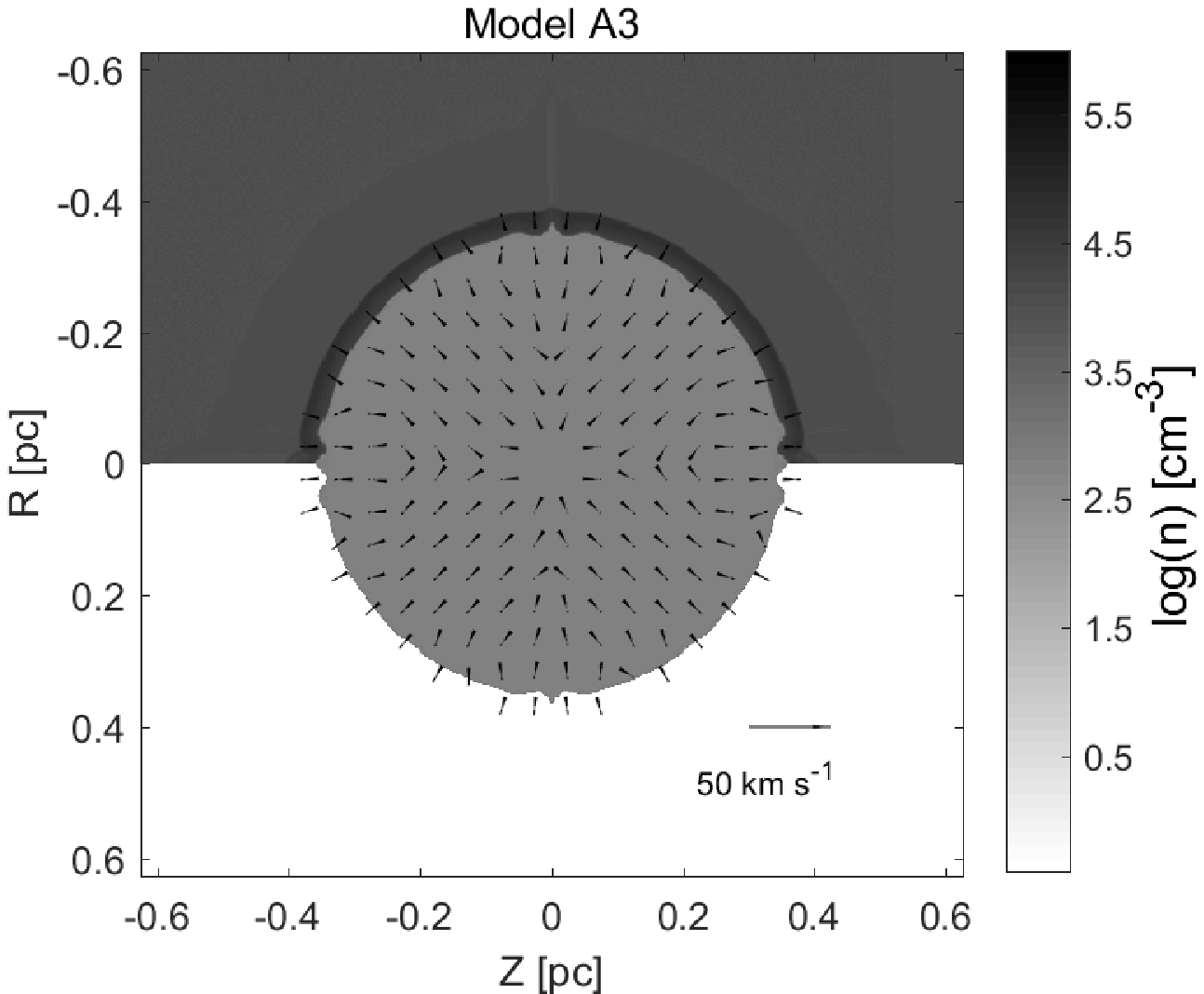}
\includegraphics[scale=0.5]{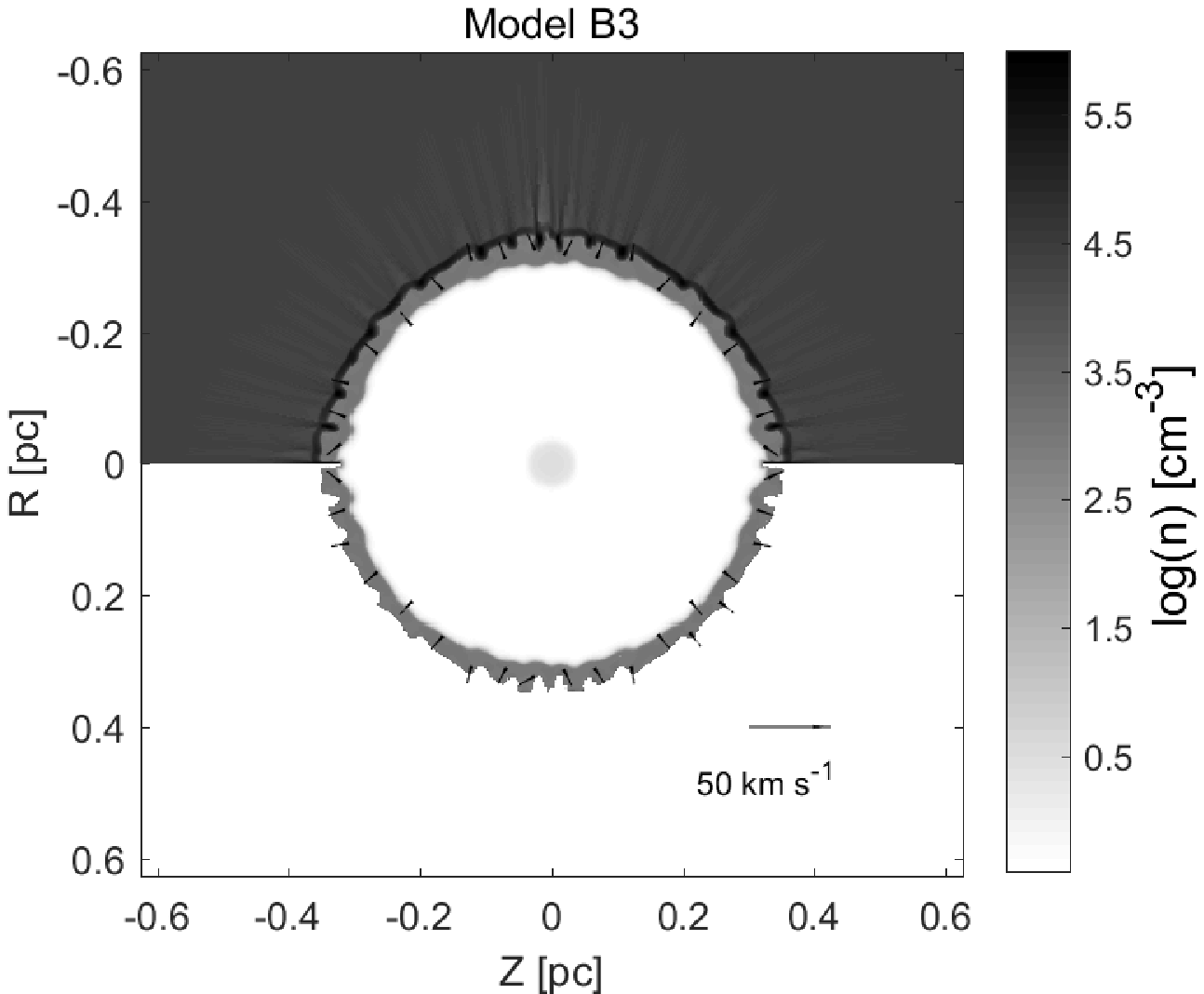}
\caption{Number density distributions of all materials and ionized hydrogen ($H^+$) in model A3 (the left panel) and B3 (the right panel) at the age of 50,000 years. The top half of the panels presents the density distribution of all materials, and that of ionized hydrogen is showed in the bottom half. The arrows show the velocity field. Only velocities higher than $1.0~km~s^{-1}$ are showed, and the velocity field in the stellar-wind bubble is not presented.}
\label{fig:modela3}
\end{figure}

In order to form this kind of H II regions, the stellar mass of the centric stars is set to be $19.0~M_\odot$. The corresponding luminosities of ionizing and dissociating radiations are both presented in Table \ref{table_masses} \citep{dia98}. The H II region evolves in uniform interstellar medium. The temperature of the ambient interstellar medium is assumed to be $T=10~K$ at the start, and the number densities are $n_H=3000,~5000,~10000,~15000,~20000~cm^{-3}$ for model A1-A5, respectively. The simulations of these models are all ceased at age of $50,000$ yr. These H II regions are all spherical as a Str$\ddot{\textrm{o}}$mgren sphere. The density distributions of all materials and ionized hydrogen in model A3 are presented in the left panel of Figure \ref{fig:modela3}. In real observations, the antenna diagram varies with the angle $\theta$ between the position and the center in the beam as $P(\theta)=exp[-ln2(2\theta/\theta_b)^2]$ with a HPBW $\theta_b$. This effect should be considered in the data reduction if the size of source is not much smaller than the beam size of the telescope. It can be solved by appropriate compensation to the beam size or using On-The-Fly mode. For simplicity, this effect is not included in calculating antenna temperature. 8 Hn$\alpha$ lines with the principal quantum numbers $n=40,~52,~63,~71,~80,~90,~100$ and $113$ from 6 bands (3-mm, Q, K, Ku, X, C) are used to estimate the electron temperatures and densities. The relative uncertainties of the observed fluxes of hydrogen recombination lines are to be same either $1\%$ or $3\%$. The average estimated values of the electron temperature and density with 1 $\sigma$ errors are listed in Table \ref{table_series1}. They are calculated by the least square method after using monte carlo method. The mass-weighted average electron temperatures and densities as well as the flux-weighted average electron temperatures and densities are also listed. In model A1-A5, the differences between the mass-weighted values and the flux-weighted values are small. The average estimated values only have small deviations from the average electron temperatures and the average electron densities. So these values are reliable estimates of the properties of the H II regions. 


\begin{table}[h]
\centering
\begin{tabular}{|c|c|c|c|c|}
\hline
Stellar Mass [$M_\odot$] & $log(L_{EUV})$ [$s^{-1}$] & $log(L_{FUV})$ [$s^{-1}$] & $\dot{M}$ [$M_\odot~yr^{-1}$]& $v_w$ [$km~s^{-1}$] \\
\hline
19.0 & 47.75 & 48.21 & 2.913$\times10^{-7}$ & 1600 \\
40.9 & 48.78 & 48.76 & 9.927$\times10^{-7}$ & 2720 \\
\hline
\end{tabular}
\caption{The luminosities of ionizing and dissociating radiations, and mass-loss rates and the terminal velocities of the stellar wind are listed for corresponding stellar masses.}\label{table_masses}
\end{table}

\begin{table}[h]
\centering
\begin{tabular}{|c|c|c|c|c|c|c|}
\hline
\multirow{2}*{Model} & \multirow{2}*{$\bar{T}_e/\bar{T}_e^f$ [K]} & \multirow{2}*{$\bar{n}_e/\bar{n}_e^f$ [$cm^{-3}$]} & \multicolumn{2}{c}{$\sigma/\mu=1\%$} & \multicolumn{2}{|c|}{$\sigma/\mu=3\%$} \\
\cline{4-7}
   &     &      & $\hat{T}$ [K]& $\hat{n}_e$ [$cm^{-3}$] & $\hat{T}$ [K]& $\hat{n}_e$ [$cm^{-3}$] \\
\hline
A1 & $11956/11833$ & $366.3/401.7$ & $12213\pm^{1123.4}_{1138.1}$ & $408.63\pm^{81.2}_{77.5}$ & $11782\pm^{3511.0}_{3277.6}$ & $429.5\pm^{246.6}_{215.7}$ \\
A2 & $12058/11996$ & $379.9/411.3$ & $12086\pm^{1120.5}_{1119.0}$ & $473.6\pm^{88.7}_{84.6}$ & $11785\pm^{3211.9}_{3196.9}$ & $517.5\pm^{258.7}_{266.3}$ \\
A3 & $12069/11980$ & $596.9/686.2$ & $12105\pm^{972.7}_{1030.5}$ & $654.7\pm^{121.6}_{117.6}$ & $11703\pm^{3003.2}_{2859.4}$ & $688.7\pm^{382.9}_{333.8}$ \\
A4 & $12080/11978$ & $719.1/856.7$ & $12124\pm^{954.1}_{940.3}$ & $797.5\pm^{157.5}_{136.8}$ & $11796\pm^{2766.8}_{2778.0}$ & $882.1\pm^{436.1}_{445.6}$ \\
A5 & $12077/11947$ & $826.7/1030.8$ & $12144\pm^{933.7}_{960.6}$ & $939.3\pm^{182.7}_{163.0}$ & $11835\pm^{2586.4}_{2728.0}$ & $1042.1\pm^{506.8}_{529.2}$ \\
\hline
\end{tabular}
\caption{The estimated electron temperatures and densities of the model A1-A5 are listed. One sigma errors of the estimated values are also presented \textbf{under} the assumption of the $1\%$ and $3\%$ uncertainties of the frequency-integrated fluxes $\int S_l d\nu$. $\bar{T}_e$ and $\bar{n}_e$ means the mass-weighted average electron temperature and the average electron density in the H II region. $\bar{T}_e^f$ and $\bar{n}_e^f$ means the flux-weighted average electron temperature and the average electron density in the H II region.}\label{table_series1}
\end{table}

\begin{figure}[ht!]
\centering
\includegraphics[scale=0.5]{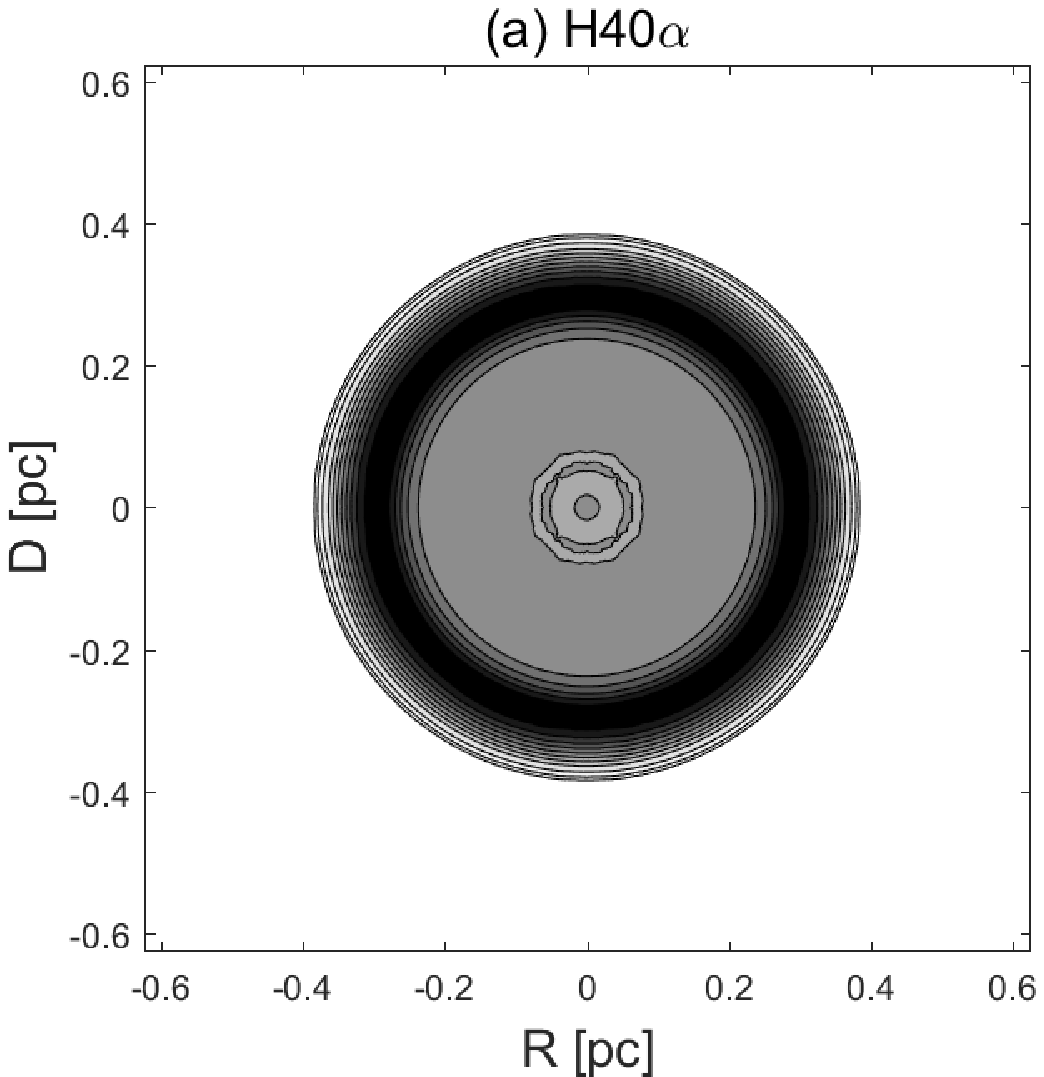}
\includegraphics[scale=0.5]{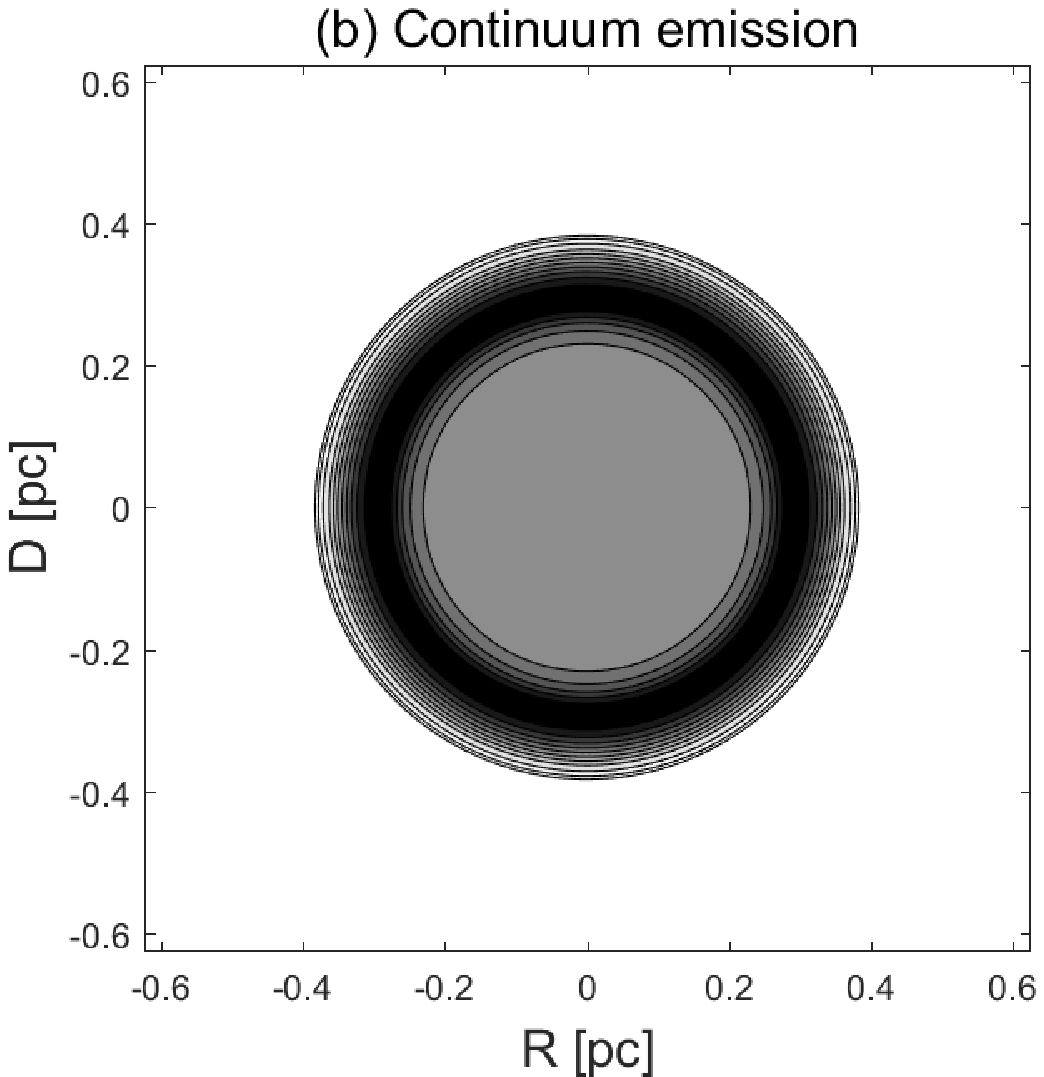}
\caption{The intensity maps of the H40$\alpha$ emission line (a) and the continuum emission (b) at the frequency of $99.023~GHz$ in model B3. The contour levels are at 3\%, 5\%, 10\%, 20\%, 30\%, 40\%, 50\%, 60\%, 70\%, 80\%, 90\% of the emission peaks in each panel.}
\label{fig:modelbimage}
\end{figure}

\begin{figure}[ht!]
\centering
\includegraphics[scale=0.3]{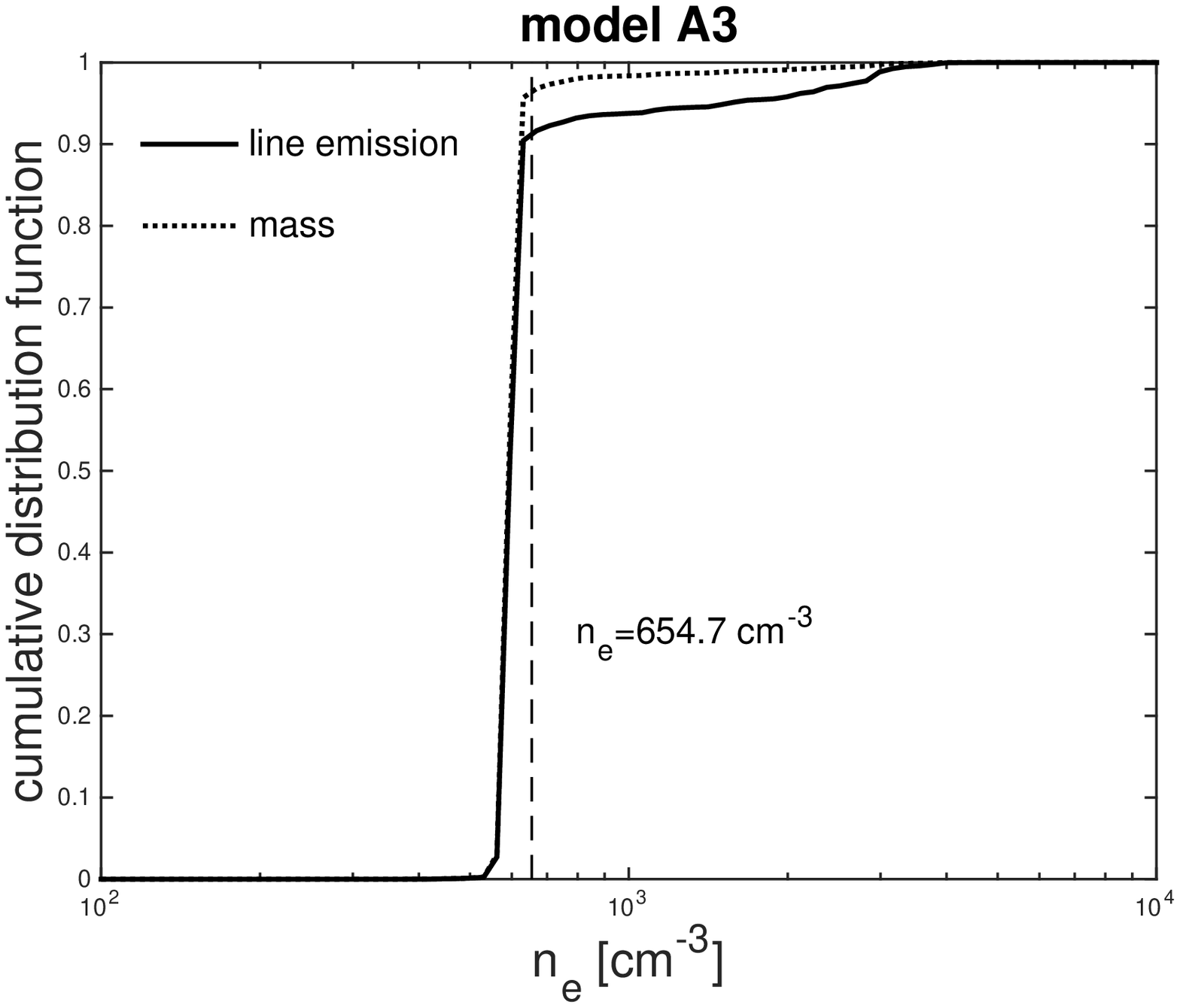}
\includegraphics[scale=0.3]{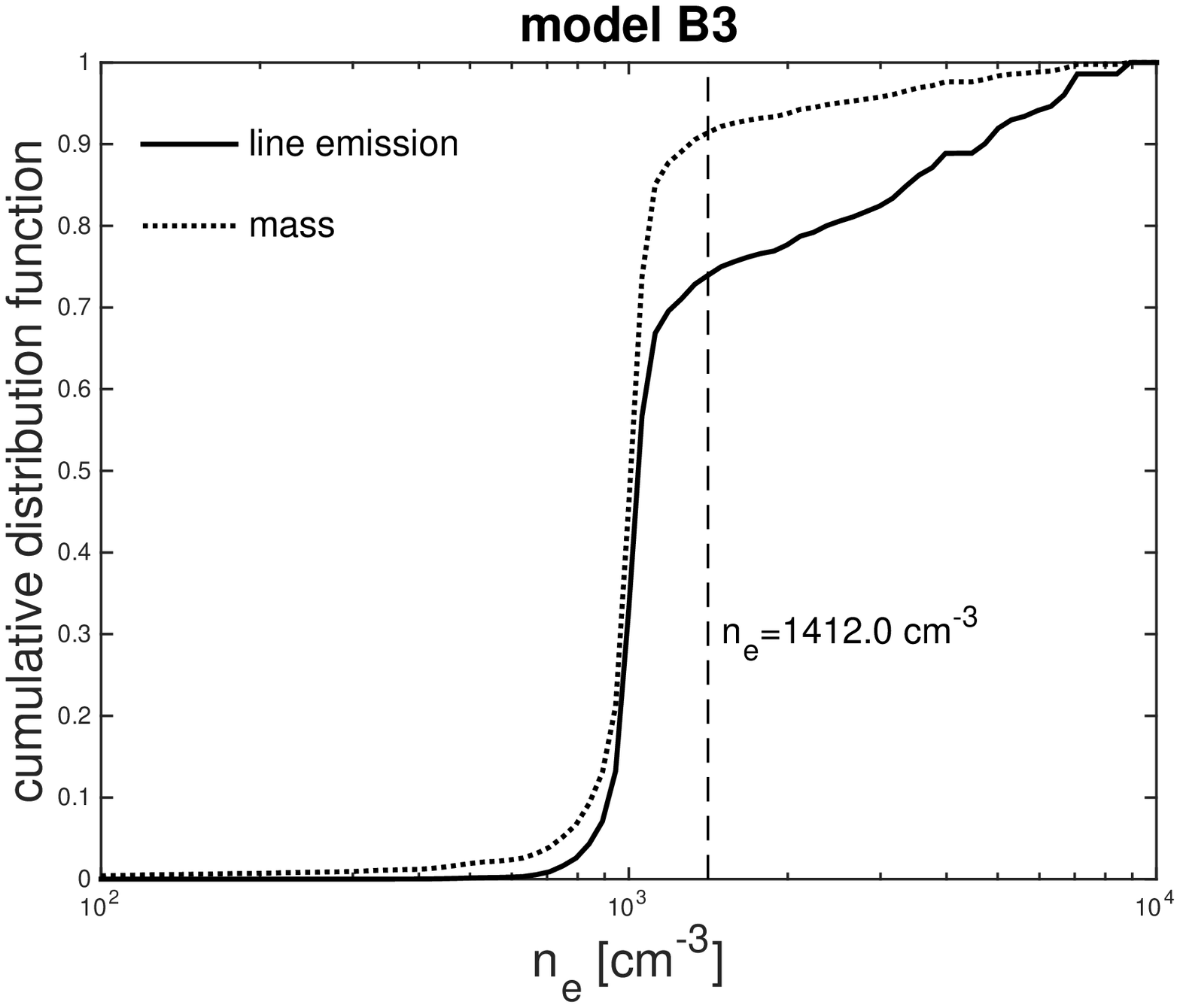}
\includegraphics[scale=0.3]{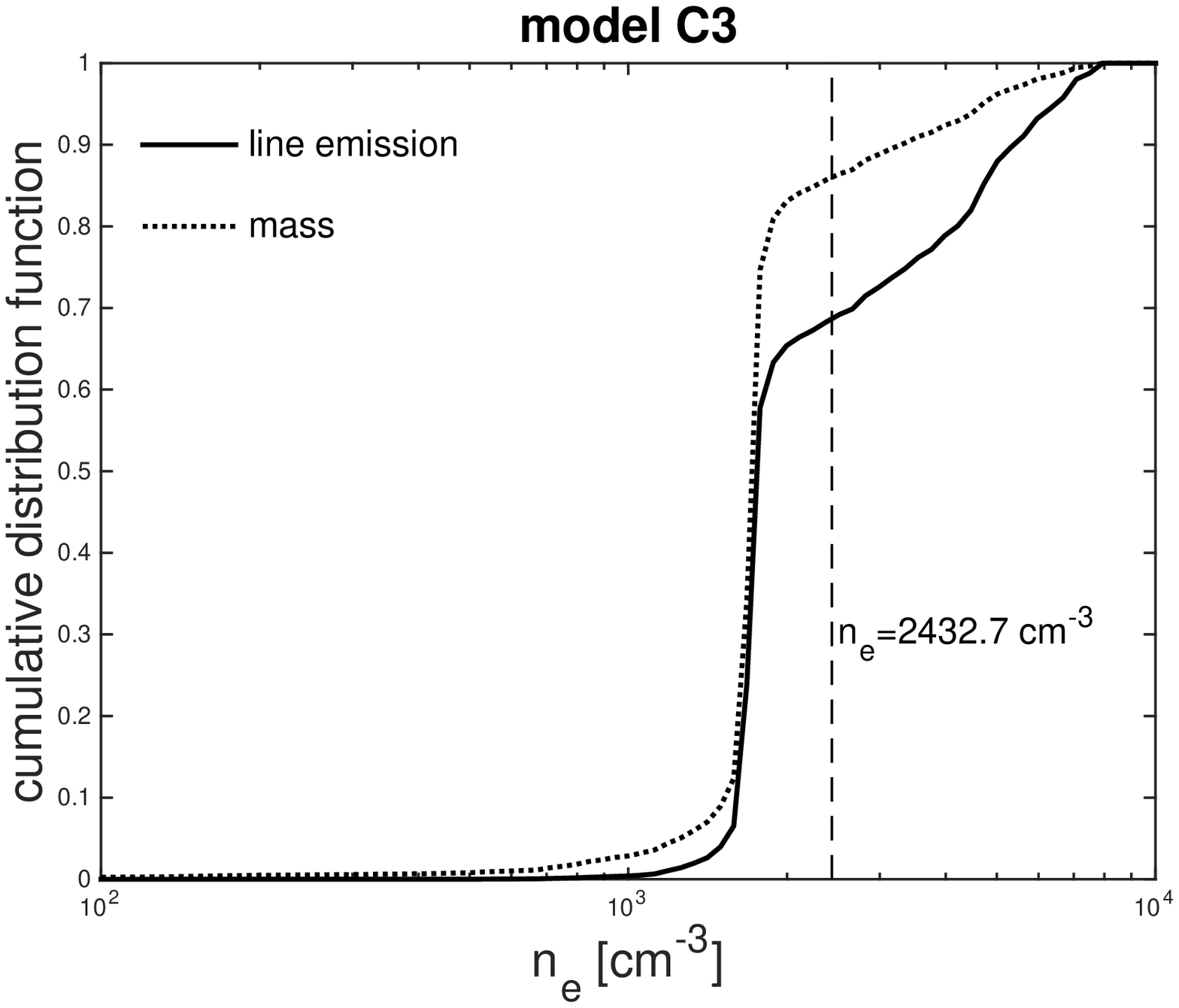}
\caption{The cumulative contribution curves of the hydrogen recombination line emission (H113$\alpha$) and the total mass of the H II region from gas in different electron density intervals for model A3 (left), B3 (middle) and C3 (right). The vertical dashed line indicates the estimated electron number density.}
\label{fig:cdfmodel1}
\end{figure}

The H II region models including a stellar wind are simulated in model B1-B5. The initial conditions are same as those in model A1-A5 except that the effect of a stellar wind is included. The properties of the stellar wind are also included in Table \ref{table_masses} \citep{dal13}. The simulations are stopped at the age of $50,000~yr$ as in model A1-A5. The density distributions in model B3 are displayed in the right panel of Figure \ref{fig:modela3}. In model B1-B5, there is a low-density and hot stellar wind bubble ($T>10^6~K$ and $n_e<1~cm^{-3}$) forming around the centric star composed of the stellar-wind material. The contribution to the recombination lines from the hot bubble are negligible due to the very low density and high temperature, and the recombination line photons are mainly emitted from the photoionized region. This is also reflected on the intensity maps of the H40$\alpha$ line and the continuum radiation at $\nu=99.023~GHz$ in Figure \ref{fig:modelbimage}. In these maps, the boundary area of the H II region is brighter than the center region. The estimated values and 1 $\sigma$ statistical errors of the electron temperatures and number densities in model B1-B5 are shown in Table \ref{table_series2}. The mass-weighted average properties in the photoionized region and flux-weighted average properties are also presented. Because the strengths of the hydrogen recombination lines are roughly proportional to $n_e^2$, the weight of high-density region is higher in flux-weighted average values than in mass-weighted average values. This is also displayed by the cumulative distribution curve of the line emission in Figure \ref{fig:cdfmodel1}.  In our simulations, it is found that the differences between the mass-weighted values and the flux-weighted values are positively related to the variations of the properties across the photoionized region. The density gradients in model B3-B5 are larger than those in model B1-B2. The estimated electron temperatures and number densities are closer to the flux-weighted average values, and are more representative of the properties of the ionized gas in the relatively high-density region. The results for model A1-A5 and model B1-B5 show that our estimation method works well for a spherical H II region whenever the effect of stellar wind is important or not.

\begin{table}[h]
\centering
\begin{tabular}{|c|c|c|c|c|c|c|}
\hline
\multirow{2}*{Model} & \multirow{2}*{$\bar{T}_e/\bar{T}_e^f$ [K]} & \multirow{2}*{$\bar{n}_e/\bar{n}_e^f$ [$cm^{-3}$]} & \multicolumn{2}{c}{$\sigma/\mu=1\%$} & \multicolumn{2}{|c|}{$\sigma/\mu=3\%$} \\
\cline{4-7}
   &     &      & $\hat{T}$ [K]& $\hat{n}_e$ [$cm^{-3}$] & $\hat{T}$ [K]& $\hat{n}_e$ [$cm^{-3}$] \\
\hline
B1 & $12049/11811$ & $445.6/547.5$ & $12027\pm^{1050.6}_{1060.3}$ & $530.7\pm^{100.3}_{94.2}$ & $11591\pm^{3406.2}_{2918.1}$ & $548.7\pm^{322.2}_{266.9}$ \\
B2 & $12169/11815$ & $550.9/718.3$ & $12083\pm^{995.3}_{1007.9}$ & $686.1\pm^{126.7}_{123.8}$ & $11787\pm^{2919.3}_{2943.2}$ & $756.9\pm^{365.2}_{385.3}$ \\
B3 & $12170/11645$ & $1015.1/1660.1$ & $12122\pm^{828.3}_{828.9}$ & $1412.0\pm^{247.6}_{263.8}$ & $11813\pm^{2467.8}_{2435.0}$ & $1597.5\pm^{801.3}_{839.9}$ \\
B4 & $12176/11703$ & $988.6/1480.6$ & $12055\pm^{895.1}_{872.0}$ & $1313.9\pm^{234.9}_{242.4}$ & $11734\pm^{2408.2}_{2447.0}$ & $1477.2\pm^{710.6}_{769.2}$ \\
B5 & $12169/11614$ & $1028.6/1711.8$ & $12102\pm^{848.6}_{808.6}$ & $1433.7\pm^{264.6}_{258.8}$ & $11778\pm^{2503.5}_{2399.3}$ & $1612.2\pm^{786.7}_{835.9}$ \\
\hline
\end{tabular}
\caption{The estimated electron temperatures and densities of the model B1-B5 are presented. One sigma errors of the estimated values are also presented \textbf{under} the assumption of the $1\%$ and $3\%$ uncertainties of the frequency-integrated luminosities $\int S_l d\nu$. $\bar{T}_e$ and $\bar{n}_e$ means the mass-weighted average electron temperature and the average electron density in the photoionized region. $\bar{T}_e^f$ and $\bar{n}_e^f$ means the flux-weighted average electron temperature and the average electron density in the H II region.}\label{table_series2}
\end{table}

The results mentioned above are based on hydrogen recombination line observation in 6 bands. This makes a high price in telescope time. A time-saving method is to observe in less bands. But reducing the number of bands would cause the decrease in the accuracy of the estimate. In order to compensate this effect, it is necessary to increase the number of the observed lines at individual band which can generally be observed simultaneously.

The values of the electron temperature and density in model A1-B5 estimated by using 16 Hn$\alpha$ lines from 3 bands (C, Ku and 3-mm) are displayed in Table \ref{table_3band1}. The results are even more accurate than those by using 8 lines from 6 bands. It is suggested that the estimated values by using lines from less bands could keep the accuracy if more recombination lines are observed.

\begin{table}[h]
\centering
\begin{tabular}{|c|c|c|c|c|c|c|}
\hline
\multirow{2}*{Model} & \multirow{2}*{$\bar{T}_e/\bar{T}_e^f$ [K]} & \multirow{2}*{$\bar{n}_e/\bar{n}_e^f$ [$cm^{-3}$]} & \multicolumn{2}{c}{$\sigma/\mu=1\%$} & \multicolumn{2}{|c|}{$\sigma/\mu=3\%$} \\
\cline{4-7}
   &     &      & $\hat{T}$ [K]& $\hat{n}_e$ [$cm^{-3}$] & $\hat{T}$ [K]& $\hat{n}_e$ [$cm^{-3}$] \\
\hline
A1 & $11956/11833$ & $366.3/401.7$ & $12003\pm^{821.7}_{928.2}$ & $400.5\pm^{46.2}_{45.7}$ & $11771\pm^{2650.9}_{2483.5}$ & $407.5\pm^{154.8}_{150.5}$ \\
A2 & $12058/11996$ & $379.9/411.3$ & $12021\pm^{803.5}_{837.5}$ & $468.1\pm^{56.7}_{60.7}$ & $11822\pm^{2459.1}_{2443.8}$ & $481.6\pm^{179.1}_{179.6}$ \\
A3 & $12069/11980$ & $596.9/686.2$ & $12028\pm^{796.1}_{845.0}$ & $647.2\pm^{94.1}_{84.9}$ & $11801\pm^{2340.7}_{2330.9}$ & $672.1\pm^{261.2}_{255.2}$ \\
A4 & $12080/11978$ & $719.1/856.7$ & $12036\pm^{788.3}_{742.8}$ & $783.8\pm^{107.4}_{107.8}$ & $11857\pm^{2284.9}_{2293.7}$ & $839.9\pm^{308.2}_{338.7}$ \\
A5 & $12077/11947$ & $826.7/1030.8$ & $12054\pm^{770.8}_{760.3}$ & $925.8\pm^{145.8}_{131.4}$ & $11845\pm^{2159.6}_{2187.4}$ & $1008.3\pm^{404.2}_{432.9}$ \\
B1 & $12049/11811$ & $445.6/547.5$ & $11885\pm^{814.8}_{810.2}$ & $523.5\pm^{65.4}_{66.4}$ & $11605\pm^{2263.1}_{2318.1}$ & $537.0\pm^{187.5}_{198.1}$ \\
B2 & $12169/11815$ & $550.9/718.3$ & $11917\pm^{782.8}_{842.2}$ & $671.4\pm^{87.2}_{82.5}$ & $11738\pm^{2266.9}_{2359.4}$ & $707.6\pm^{269.7}_{271.0}$ \\
B3 & $12170/11645$ & $1015.1/1660.1$ & $11916\pm^{659.9}_{732.7}$ & $1366.2\pm^{218.7}_{218.1}$ & $11652\pm^{1947.5}_{1994.6}$ & $1599.1\pm^{639.7}_{786.2}$ \\
B4 & $12176/11703$ & $988.6/1480.6$ & $11889\pm^{687.0}_{705.6}$ & $1276.8\pm^{202.3}_{205.3}$ & $11635\pm^{1832.7}_{1977.0}$ & $1456.9\pm^{584.8}_{698.3}$ \\
B5 & $12169/11614$ & $1028.6/1711.8$ & $11908\pm^{668.4}_{724.2}$ & $1383.5\pm^{238.3}_{235.4}$ & $11628\pm^{1839.3}_{1970.4}$ & $1616.7\pm^{674.2}_{803.9}$ \\
\hline
\end{tabular}
\caption{The estimated electron temperatures and densities of the model A1-B5 are presented. The properties of the H II regions are estimated by using 16 Hn$\alpha$ lines from 3 bands.}\label{table_3band1}
\end{table}

\subsection{properties of a spherical H II region with $EM>1.0\times10^5~cm^{-6}pc$} \label{sec:solid}

The radio continuum optical depths in the above models are relatively low. However, a large number of H II regions in the Galaxy that we can observe do not have such low continuum optical depths. In the model C1-C5, the evolutions of spherical H II regions with a $40.9~M_\odot$ star are simulated. The stellar parameters consistent with a $40.9~M_\odot$ star are presented in Table \ref{table_masses}. The other initial conditions in models C1-C5 are all the same as those in models B1-B5, and these simulations are also ceased at $50,000~yr$. Although the stellar mass is higher, the structure of the H II regions in model C1-C5 is not essentially different from that in model B1-B5. The emissions of the radio hydrogen recombination lines still mainly come from the photoionized region and represent the properties of the ionized gas in this region. In these models, the parameter $b_m(1-\tau_{\nu,C}\beta/2)$ is no longer approximately equal to the departure coefficient $b_m$. Since the LOS depth become a new free parameter that needs to be fit, it is necessary to use more lines in order to keep the accuracy of the estimate as in model A1-B5. The estimated values and errors in model C1-C5 are calculated from 16 Hn$\alpha$ lines in 6 bands and listed in Table \ref{table_series3no}. In model C1-C5, the differences between the mass-weighted and flux-weighted average values are higher than those in model B1-B5. That means a larger variation of the density across the photoionized region, and is reflected on the cumulative distribution curves presented in the right panel of Figure \ref{fig:cdfmodel1}. By the cumulative distribution curves, it is obviously showed that the estimated values are more affected by the properties in the high-density region. 


\begin{figure}[ht!]
\centering
\includegraphics[scale=0.5]{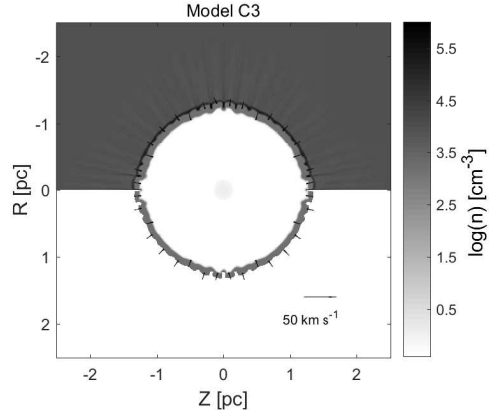}
\caption{Number density distributions of all materials and ionized hydrogen ($H^+$) in model C3 as at the age of 50,000 years. The top half of the figure presents the density distribution of all materials, and that of ionized hydrogen is showed in the bottom half. The arrows show the velocity field. Only velocities higher than $1.0~km~s^{-1}$ are showed, and the velocity field in the stellar-wind bubble is not presented.}
\label{fig:modelc3}
\end{figure}



\begin{table}[h]
\centering
\begin{tabular}{|c|c|c|c|c|c|c|}
\hline
\multirow{2}*{Model} & \multirow{2}*{$\bar{T}_e/\bar{T}_e^f$ [K]} & \multirow{2}*{$\bar{n}_e/\bar{n}_e^f$ [$cm^{-3}$]} & \multicolumn{2}{c}{$\sigma/\mu=1\%$} & \multicolumn{2}{|c|}{$\sigma/\mu=3\%$} \\
\cline{4-7}
   &     &      & $\hat{T}$ [K]& $\hat{n}_e$ [$cm^{-3}$] & $\hat{T}$ [K]& $\hat{n}_e$ [$cm^{-3}$] \\
\hline
C1 & $12206/11982$ & $1008.9/1211.3$ & $12508\pm^{1496.8}_{1540.8}$ & $1202.3\pm^{210.3}_{225.0}$ & $13119\pm^{4418.4}_{4614.2}$ & $1489.1\pm^{909.7}_{764.7}$ \\
C2 & $12227/11949$ & $1275.9/1706.1$ & $12488\pm^{1653.7}_{1628.3}$ & $1622.0\pm^{373.3}_{333.8}$ & $13413\pm^{4215.9}_{4124.6}$ & $2161.1\pm^{1306.3}_{1206.1}$ \\
C3 & $12289/11915$ & $1759.3/2567.7$ & $12601\pm^{1266.8}_{1197.0}$ & $2432.7\pm^{518.5}_{527.2}$ & $12989\pm^{4041.2}_{3701.9}$ & $3112.5\pm^{2257.8}_{1763.5}$ \\
C4 & $12214/11796$ & $2023.5/3734.5$ & $12831\pm^{1310.7}_{1314.9}$ & $2938.3\pm^{609.9}_{647.4}$ & $12621\pm^{3596.5}_{3424.2}$ & $3592.1\pm^{2433.5}_{2078.5}$ \\
C5 & $12162/11616$ & $2215.3/6088.6$ & $12950\pm^{1192.2}_{1206.0}$ & $3857.1\pm^{929.2}_{905.9}$ & $12918\pm^{3458.5}_{3165.8}$ & $4987.9\pm^{3924.6}_{2898.7}$ \\
\hline
\end{tabular}
\caption{The estimated electron temperatures and densities of the model C1-C5 are presented. One sigma errors of the estimated values are also presented \textbf{under} the assumption of the $1\%$ and $3\%$ uncertainties of the frequency-integrated luminosities $\int S_l d\nu$. $\bar{T}_e$ and $\bar{n}_e$ means the mass-weighted average electron temperature and the average electron density in the photoionized region. $\bar{T}_e^f$ and $\bar{n}_e^f$ means the flux-weighted average electron temperature and the average electron density in the H II region.}\label{table_series3no}
\end{table}

The estimated values by using 18 Hn$\alpha$ lines from 4 bands are also computed and written in Table \ref{table_3band3}. The estimated values are still good reference values for the properties of the H II regions.

\begin{table}[h]
\centering
\begin{tabular}{|c|c|c|c|c|c|c|}
\hline
\multirow{2}*{Model} & \multirow{2}*{$\bar{T}_e/\bar{T}_e^f$ [K]} & \multirow{2}*{$\bar{n}_e/\bar{n}_e^f$ [$cm^{-3}$]} & \multicolumn{2}{c}{$\sigma/\mu=1\%$} & \multicolumn{2}{|c|}{$\sigma/\mu=3\%$} \\
\cline{4-7}
   &     &      & $\hat{T}$ [K]& $\hat{n}_e$ [$cm^{-3}$] & $\hat{T}$ [K]& $\hat{n}_e$ [$cm^{-3}$] \\
\hline
C1 & $12206/11982$ & $1008.9/1211.3$ & $12559\pm^{1309.5}_{1484.0}$ & $1209.2\pm^{203.4}_{209.2}$ & $13435\pm^{4624.1}_{4416.7}$ & $1515.1\pm^{775.8}_{756.5}$ \\
C2 & $12227/11949$ & $1275.9/1706.1$ & $12646\pm^{1358.3}_{1352.8}$ & $1637.7\pm^{267.8}_{288.7}$ & $13102\pm^{4264.6}_{3905.2}$ & $1995.9\pm^{1094.4}_{1018.7}$ \\
C3 & $12289/11915$ & $1759.3/2567.7$ & $12599\pm^{1133.9}_{1305.9}$ & $2442.2\pm^{509.1}_{492.3}$ & $12818\pm^{3558.2}_{3254.9}$ & $2952.8\pm^{1514.0}_{1404.0}$ \\
C4 & $12214/11796$ & $2023.5/3734.5$ & $12703\pm^{1029.9}_{1186.8}$ & $2864.1\pm^{524.3}_{519.9}$ & $12845\pm^{3214.5}_{3281.4}$ & $3539.2\pm^{2084.2}_{1841.0}$ \\
C5 & $12162/11616$ & $2215.3/6088.6$ & $13094\pm^{1187.0}_{1118.2}$ & $3852.7\pm^{824.6}_{832.8}$ & $12687\pm^{2756.4}_{2934.5}$ & $4368.9\pm^{2875.5}_{2181.1}$ \\
\hline
\end{tabular}
\caption{The estimated electron temperatures and densities of the model C1-C5 are presented. The properties of the H II regions are estimated by using 18 Hn$\alpha$ lines from 4 bands}\label{table_3band3}
\end{table}

From the results listed in Table \ref{table_series3no} and \ref{table_3band3}, it is suggested that the electron temperature and density of a spherical H II region can be estimated with relative errors $\sigma/\mu<13\%$ for $T_e$ and $<25\%$ for $n_e$ by using the strengths of multiple hydrogen recombination lines if the observational uncertainties of the frequency-integrated fluxes $\int S_ld \nu$ are lower than $1\%$. And if $\int S_ld \nu<3\%$, the errors are $\sigma/\mu<33\%$ and $<65\%$ for $T_e$ and $n_e$, respectively.

\subsection{non-spherical H II region models}


After studying performance of the estimation method on spherical H II region models, we apply this method to three non-spherical H II region models and assess the results. Since the symmetry of the morphology is reduced further, it is expected that the average effect plays even more important role in the final result. 

\begin{figure}[ht!]
\centering
\includegraphics[scale=0.4]{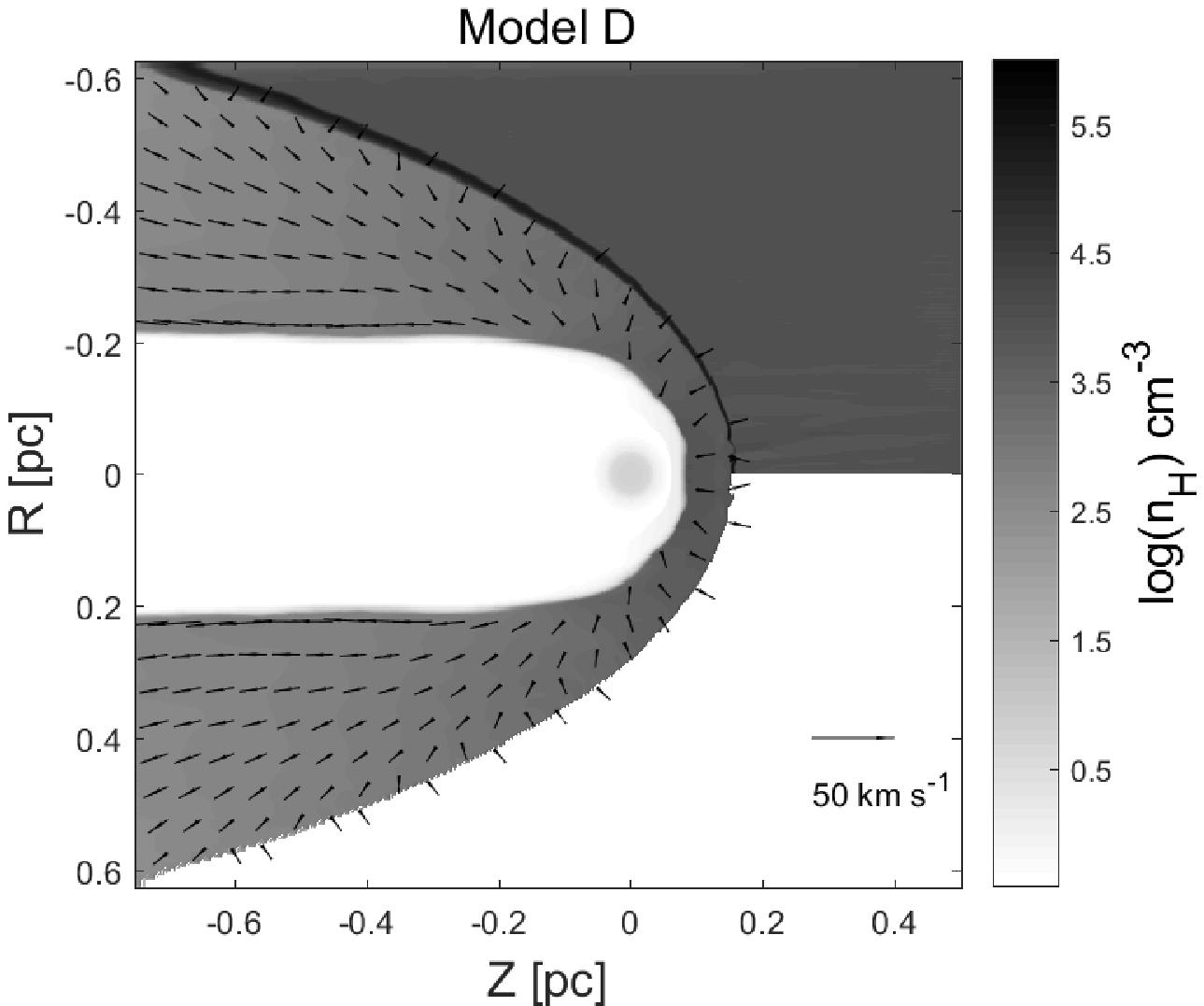}
\includegraphics[scale=0.4]{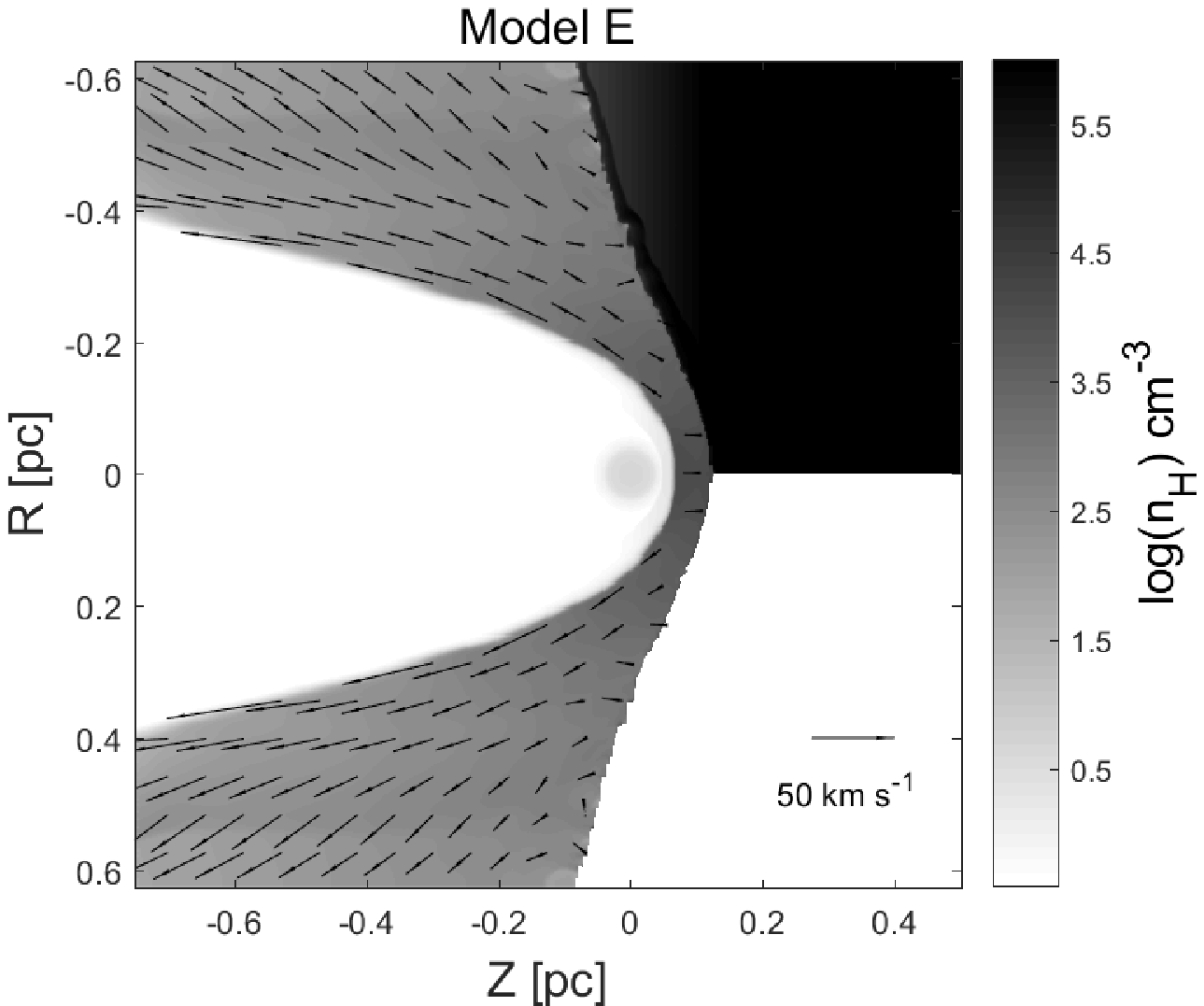}
\includegraphics[scale=0.4]{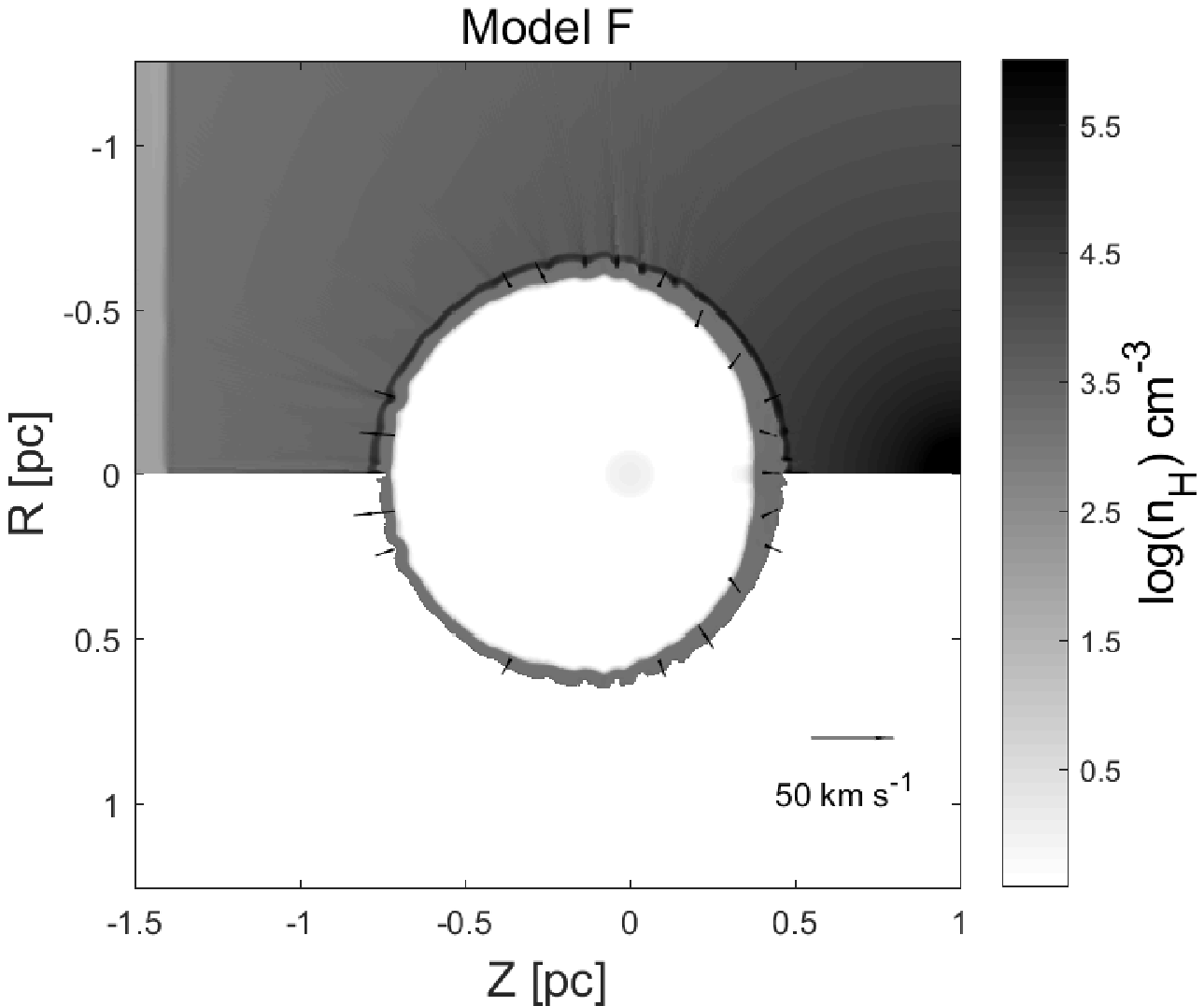}
\caption{Number density distributions of all materials and ionized hydrogen ($H^+$) in model D (left), E (middle) and F (right) at the terminal times. The top half of the panels presents the density distribution of all materials, and that of ionized hydrogen is showed in the bottom half. }
\label{fig:modseries4}
\end{figure}

\begin{figure}[ht!]
\centering
\includegraphics[scale=0.4]{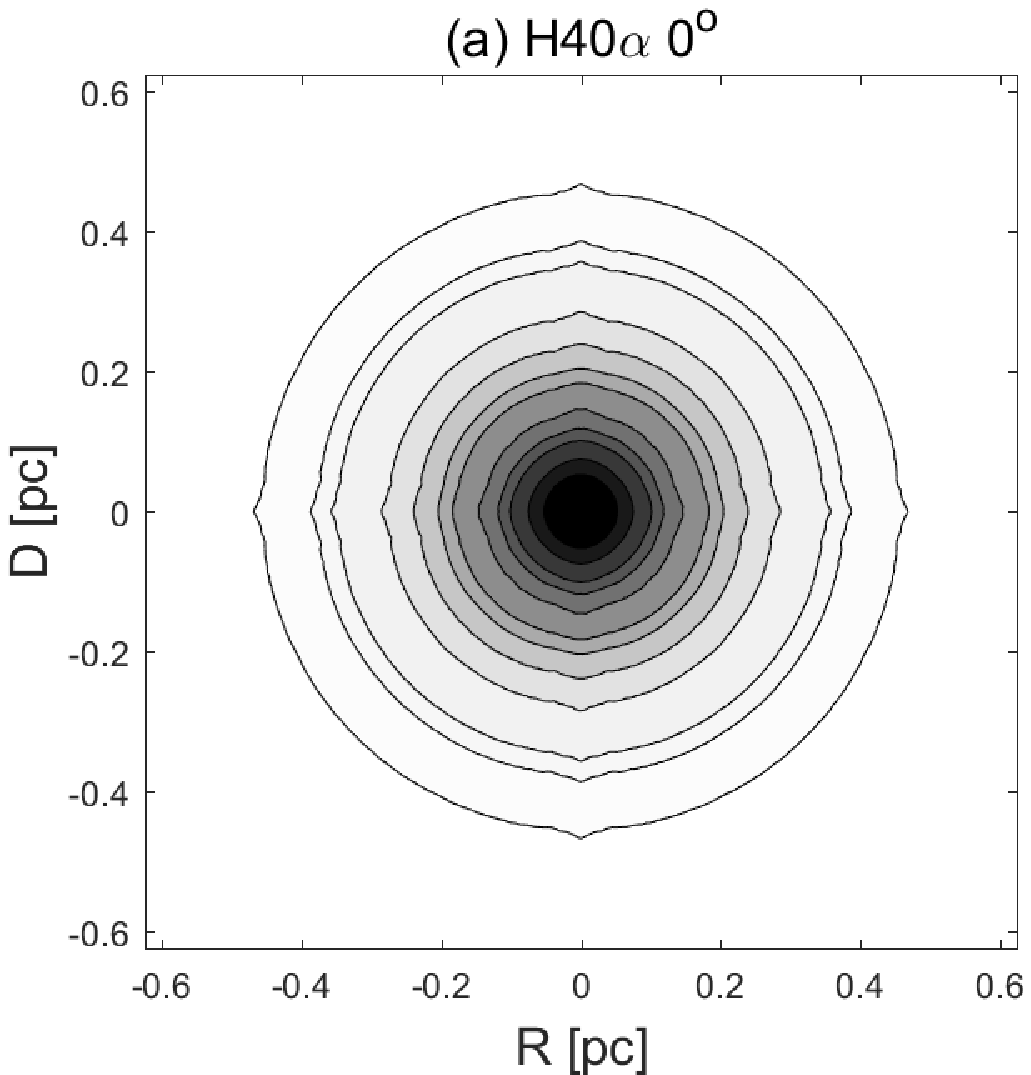}
\includegraphics[scale=0.4]{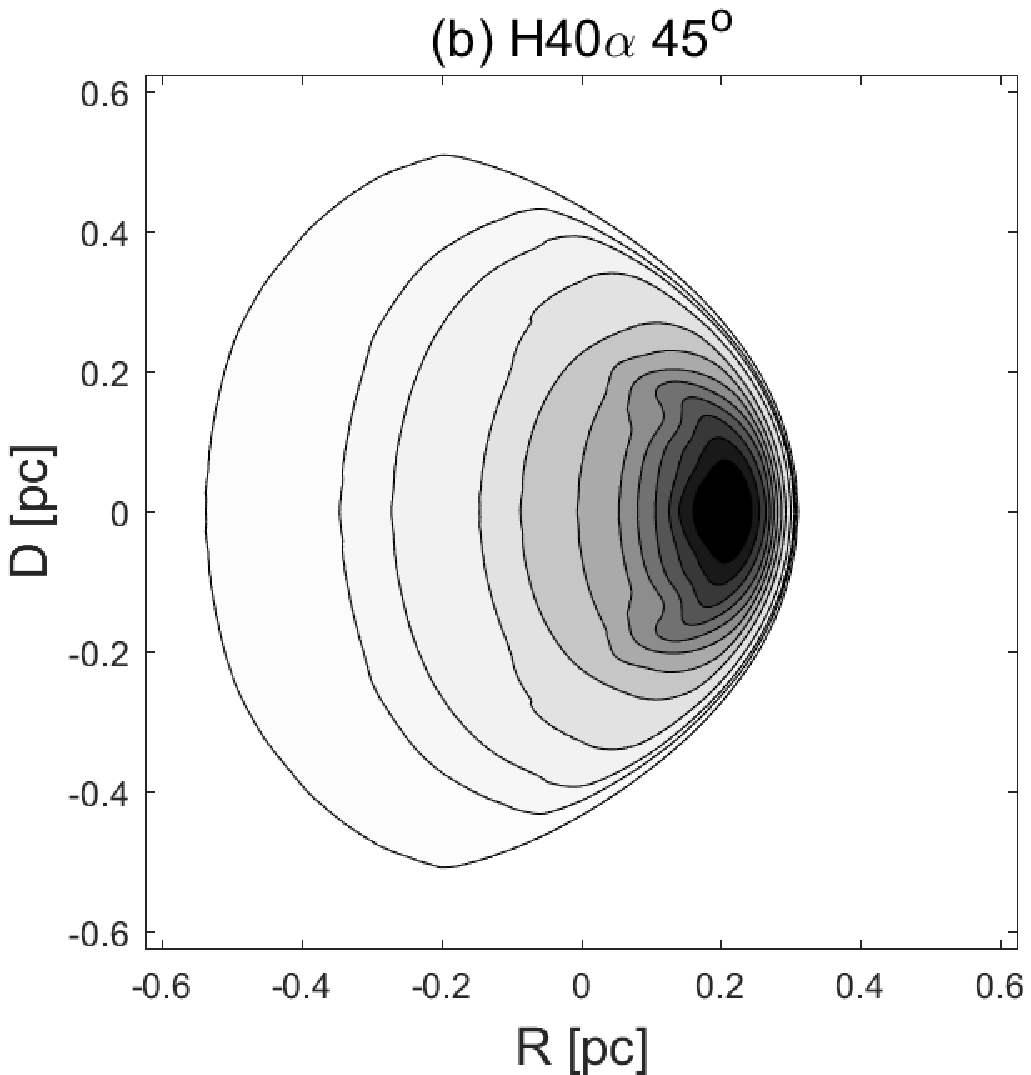}
\includegraphics[scale=0.4]{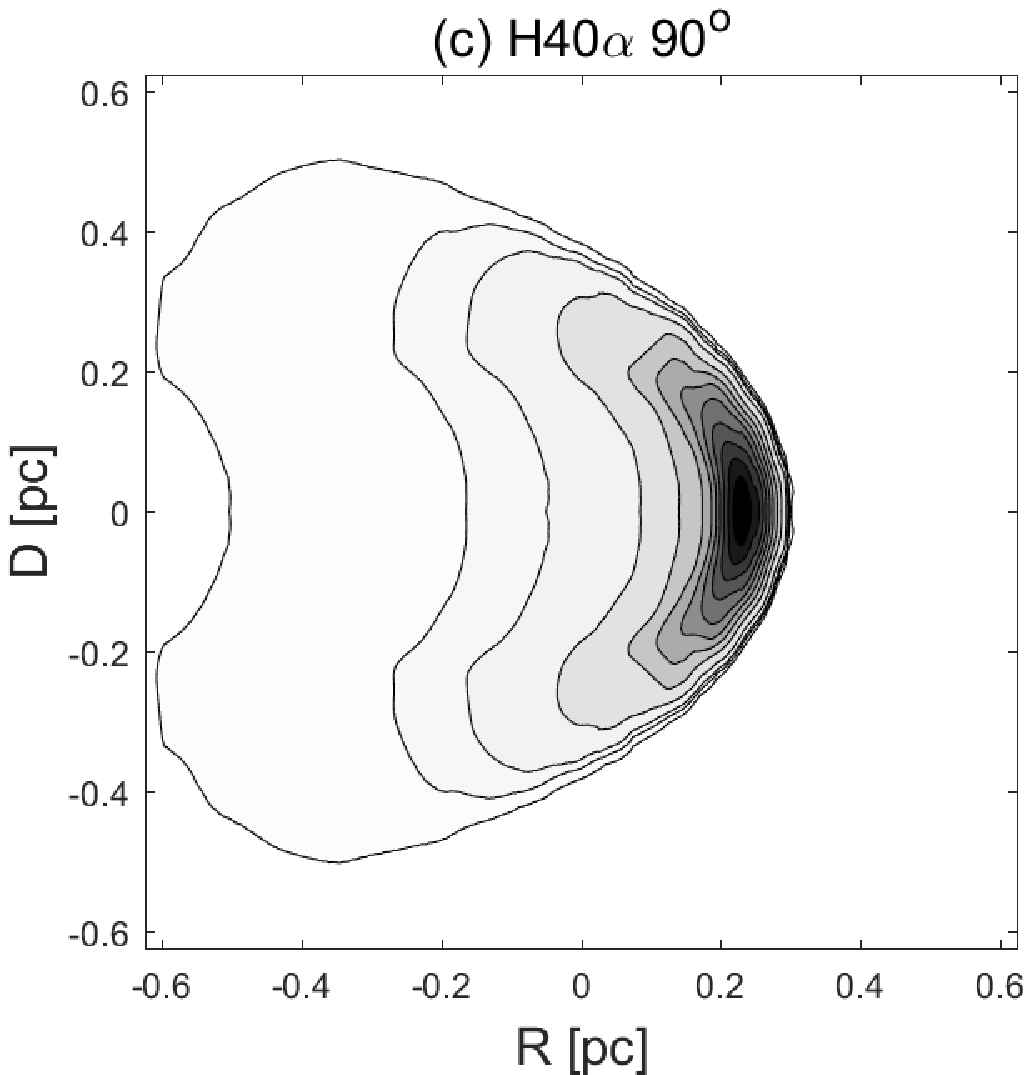}
\includegraphics[scale=0.4]{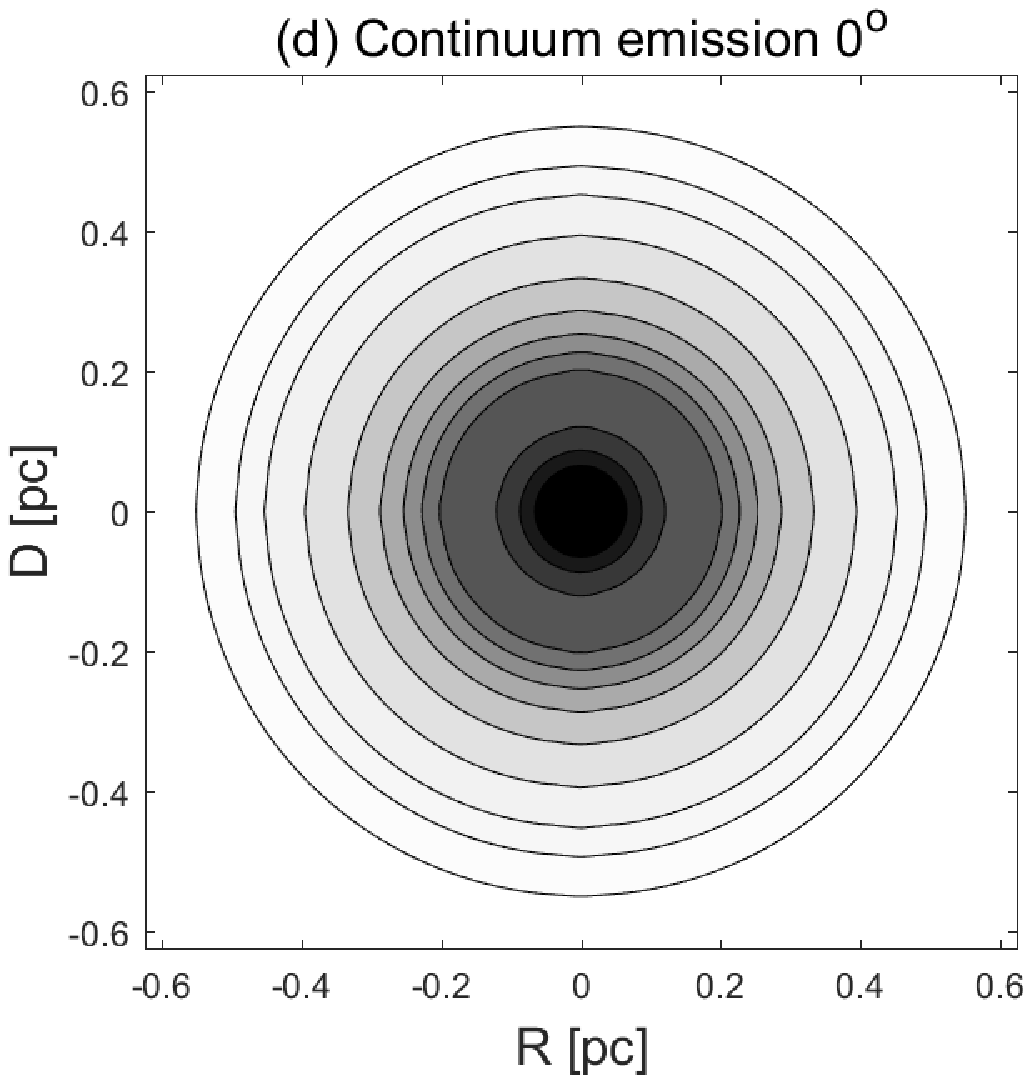}
\includegraphics[scale=0.4]{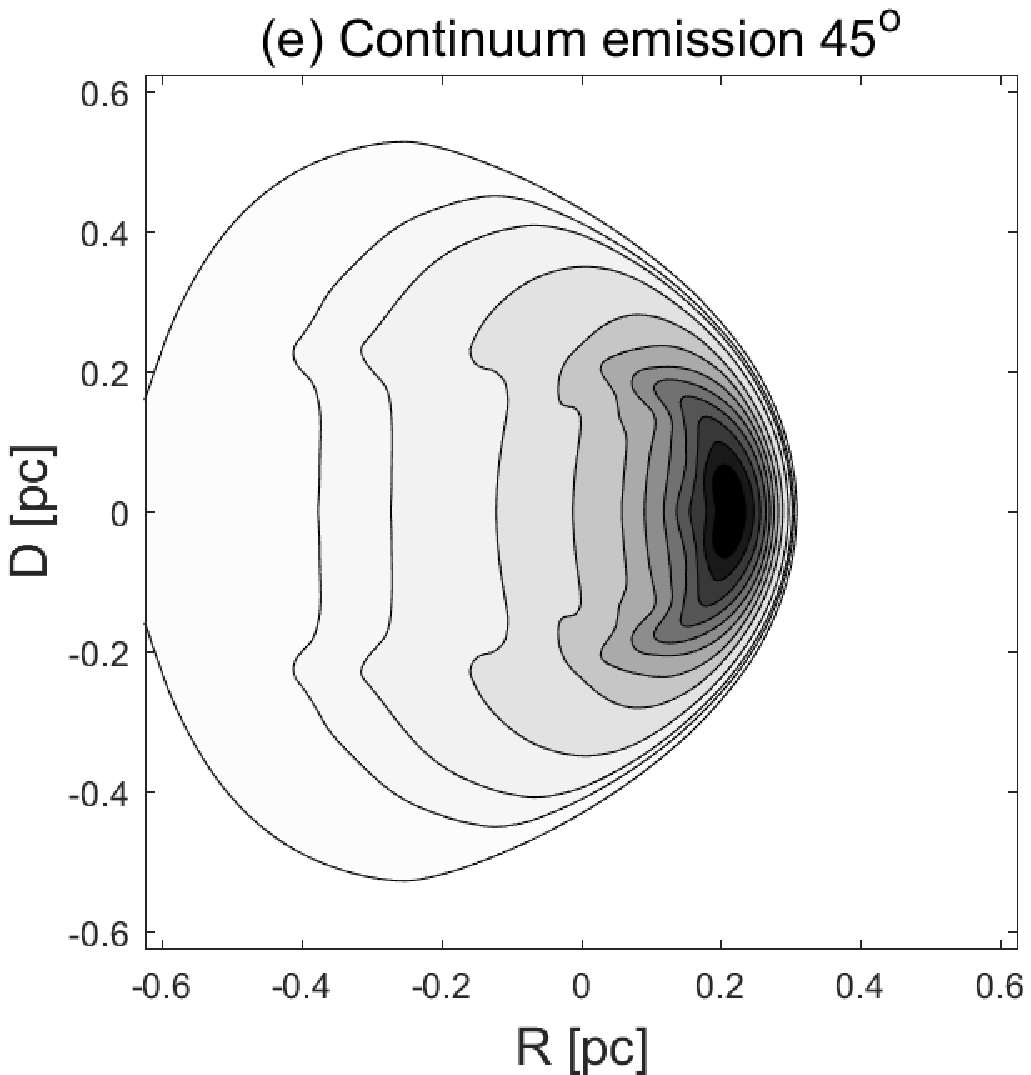}
\includegraphics[scale=0.4]{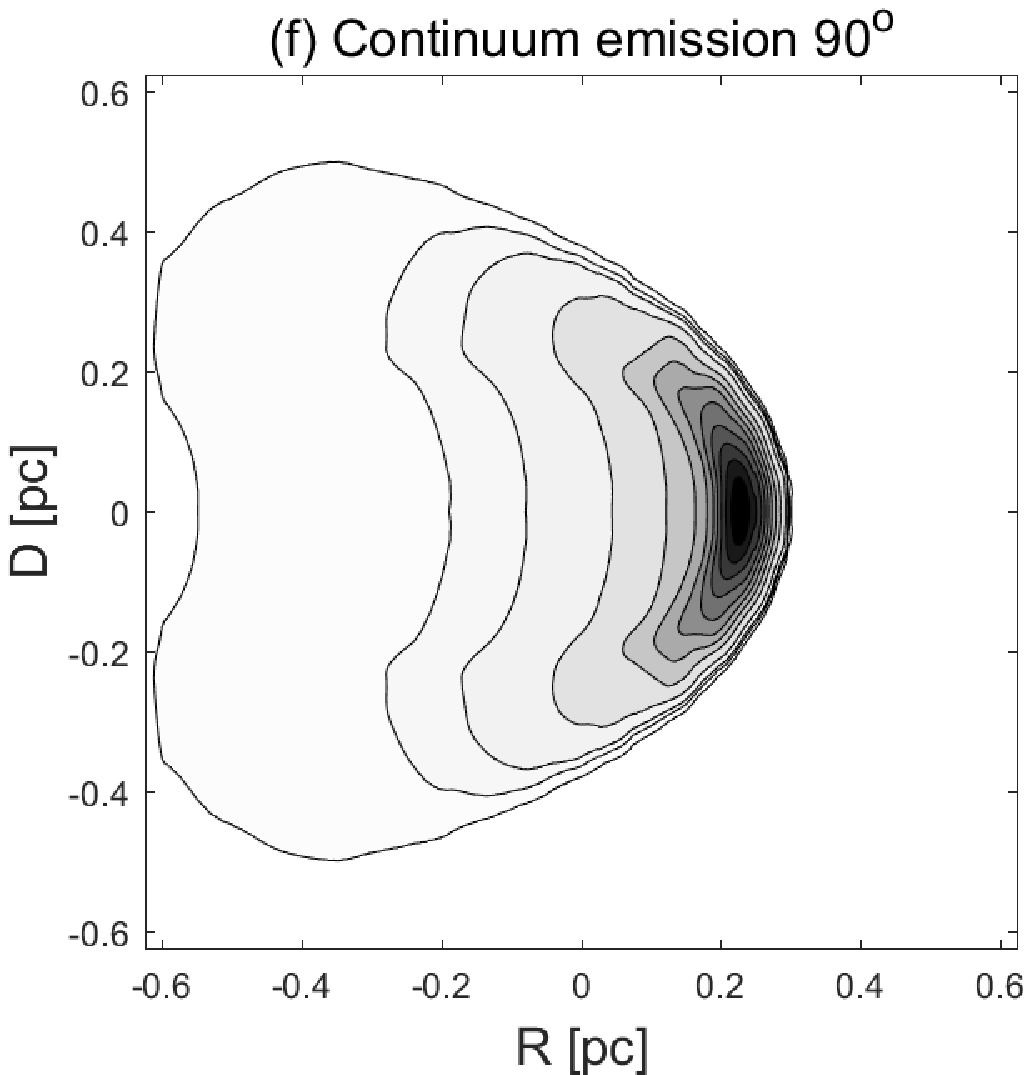}
\caption{The intensity maps of the H40$\alpha$ emission line (a-c) and the continuum emission (d-e) at the frequency of $99.023~GHz$ ($\theta=0^o,~45^o,~90^o$) in model D. The contour levels are at 1\%, 3\%, 5\%, 10\%, 20\%, 30\%, 40\%, 50\%, 60\%, 70\%, 80\%, 90\% of the emission peaks in each panel.}
\label{fig:modeimage}
\end{figure}

In model D, a bow shock in the uniform medium is simulated. The stellar mass is assumed to be $40.9~M_\odot$. The number density and the temperature of the neutral ambient medium is $n_H=8000~cm^{-3}$ and $T=10~K$, respectively. The stellar velocity is $v_*=15~km~s^{-1}$. The simulation is stopped at the age of $100,000$ years. The density distributions of all materials and ionized gas are displayed in Figure \ref{fig:modseries4}. Different from spherical H II region models, in model D the density gradient in photoionized region is much more significant. The electron density decreases from $\sim10,000~cm^{-3}$ near the apex to $400~cm^{-3}$ in the tail. The temperature is about $12000~K$ and varies slightly in the photoionized region. The intensity maps of the H40$\alpha$ line and the continuum radiation at $\nu=99.023~GHz$ for different inclination angles in model D are presented in Figure \ref{fig:modeimage}. It is obvious that the region near the apex is brightest in the images. The fractions of the hydrogen recombination line emissions contributed from the head region ($Z>0~pc$) are about $50\%$. The estimated electron temperature and density with their statistical errors are written in Table \ref{table_series4}. The observed direction is parallel to the symmetric axis from the negative side to the positive side. The estimated electron number density is more representative of the ionized gas in relatively high-density region due that the emissivity of the recombination lines are proportional to $n_e^2$. This makes the estimated electron density obviously higher than the average density in the photoionized region. The estimated temperature is about 2000 K higher than the mass-weighted and flux-weighted average temperatures. \textbf{This estimated deviation is because of the large density gradient.}

\begin{figure}[ht!]
\centering
\includegraphics[scale=0.4]{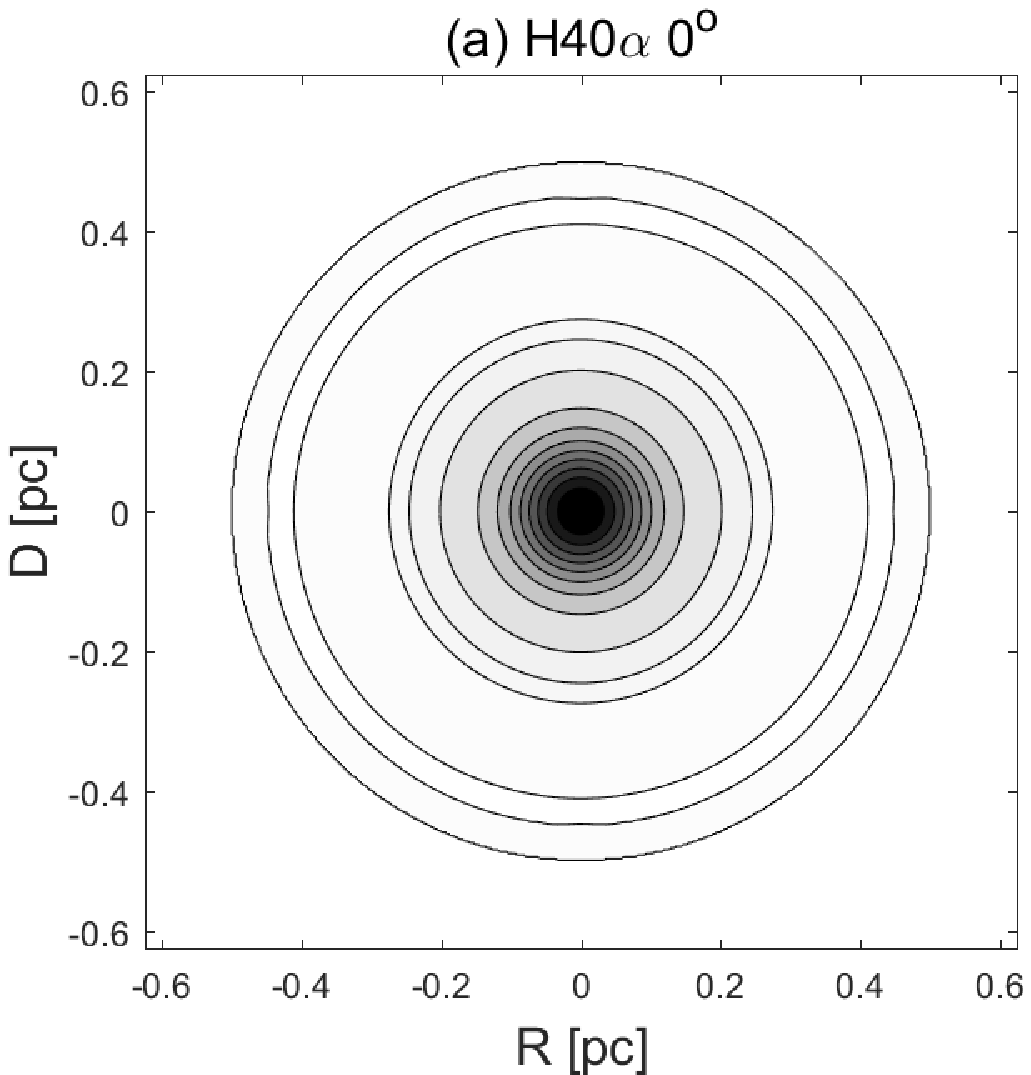}
\includegraphics[scale=0.4]{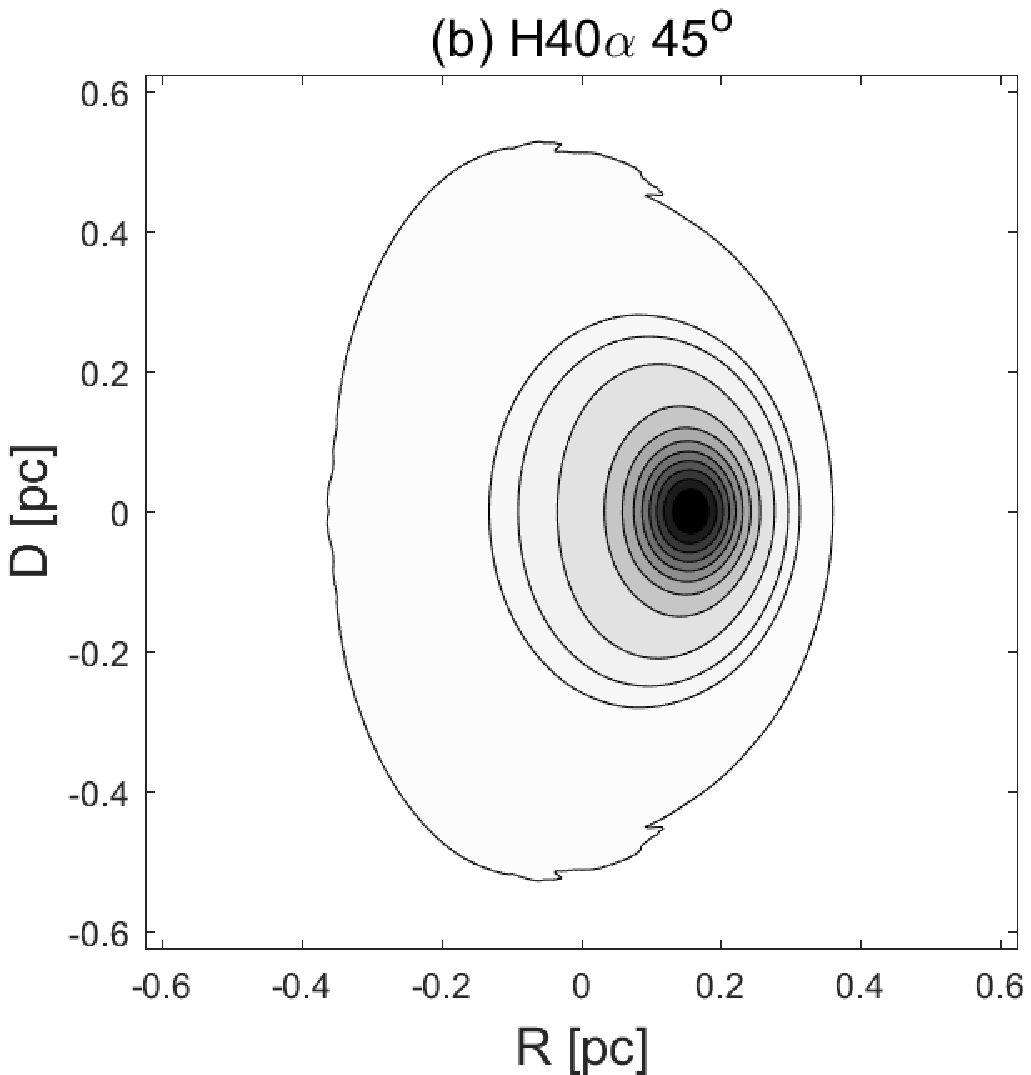}
\includegraphics[scale=0.4]{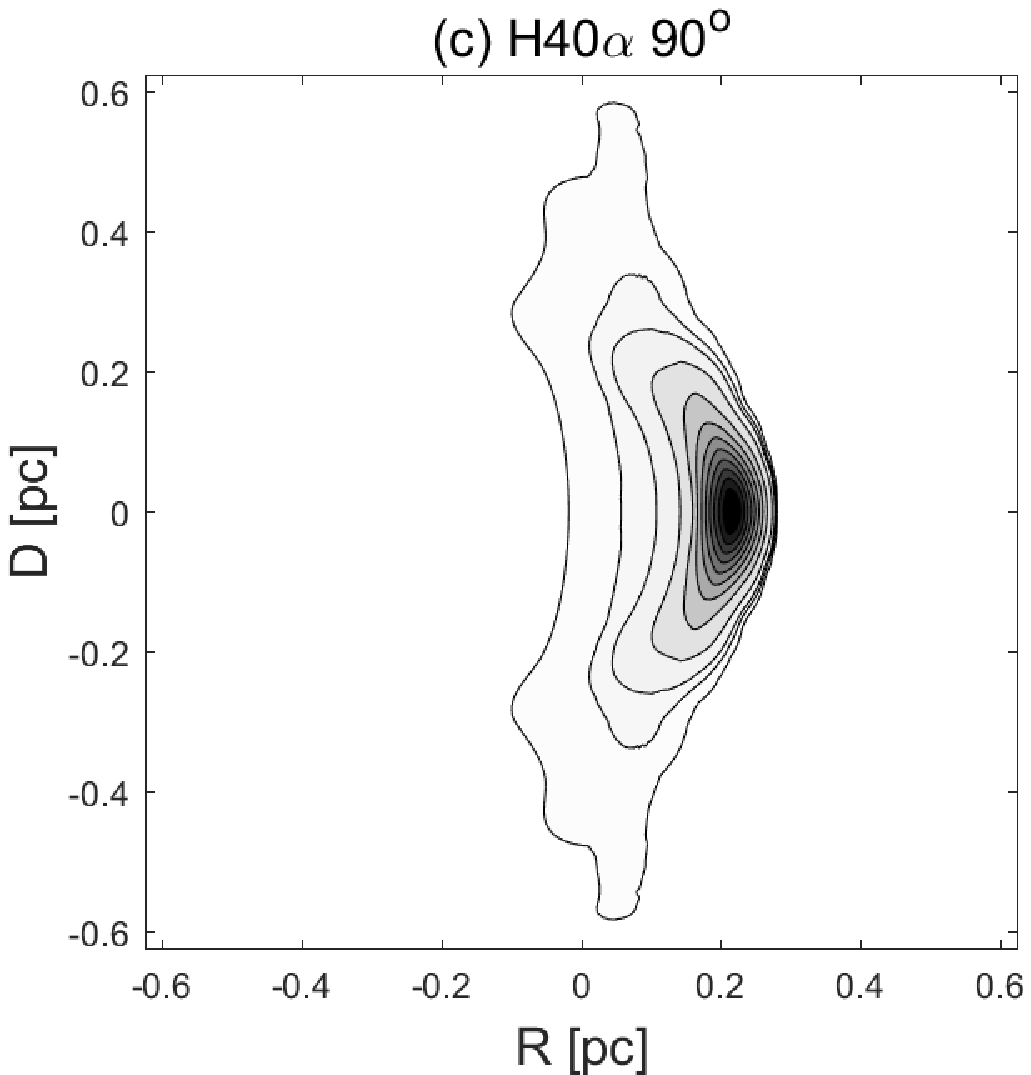}
\includegraphics[scale=0.4]{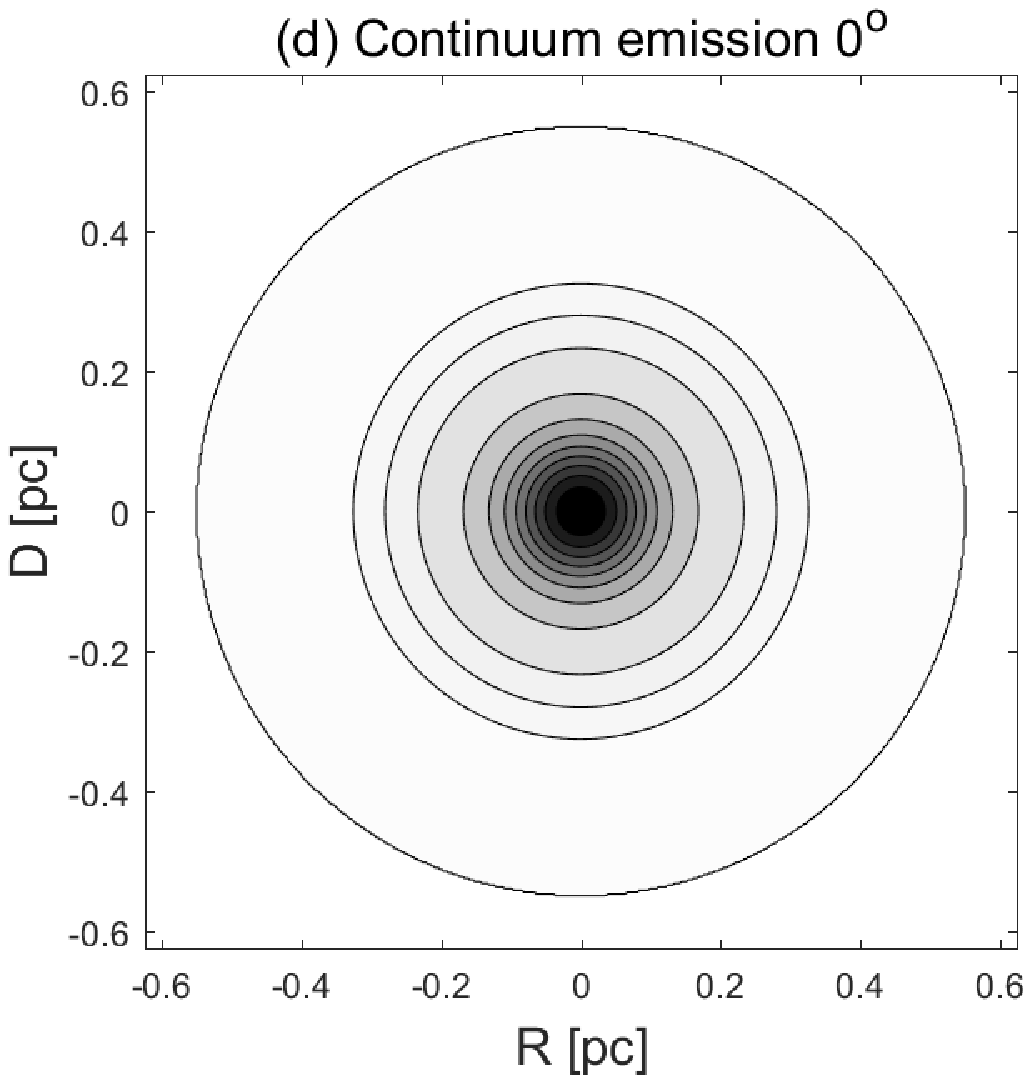}
\includegraphics[scale=0.4]{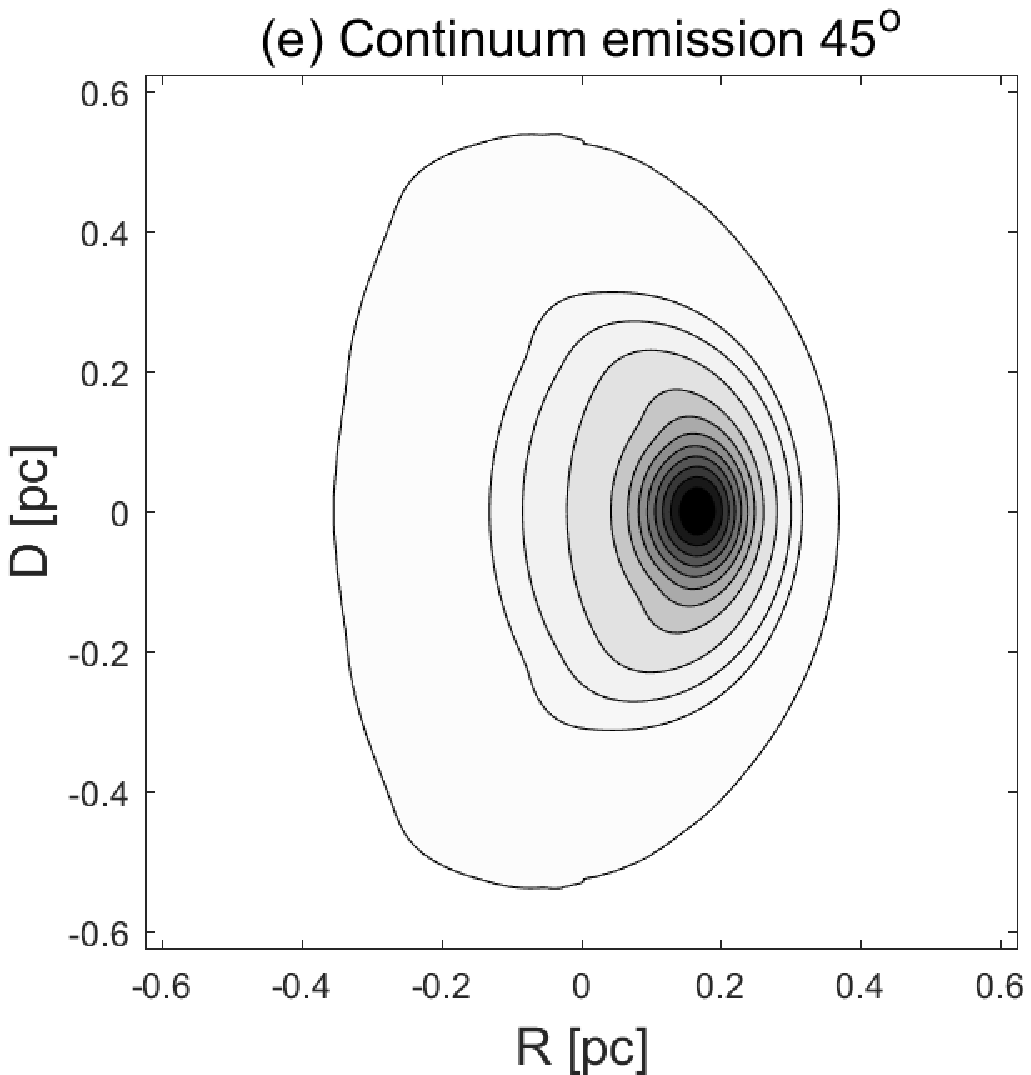}
\includegraphics[scale=0.4]{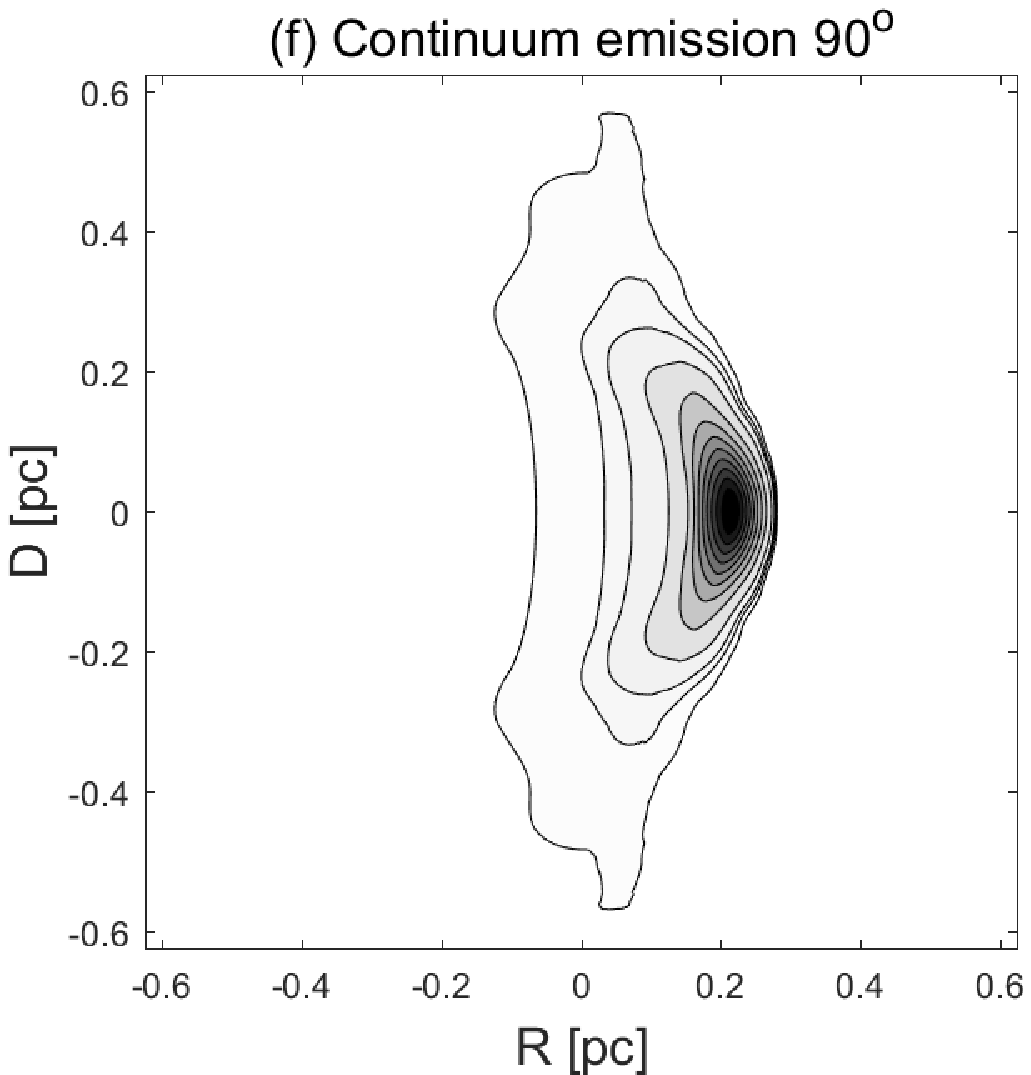}
\caption{The intensity maps of the H40$\alpha$ emission line (a-c) and the continuum emission (d-e) at the frequency of $99.023~GHz$ ($\theta=0^o,~45^o,~90^o$) in model E. The contour levels are at 1\%, 3\%, 5\%, 10\%, 20\%, 30\%, 40\%, 50\%, 60\%, 70\%, 80\%, 90\% of the emission peaks in each panel.}
\label{fig:modfimage}
\end{figure}

In model E, the evolution of a champagne-flow model is simulated. The density of the ambient medium is supposed to follow an exponential law as $n_H(z)=n_0exp(z/H)$, where $z$ is the coordinate pointing from the surface to the steep of the parental cloud. The number density $n_0$ is assumed to be $50,000~cm^{-3}$, and the scale height $H$ is $0.05~pc$. There is no stellar velocity in model E. The stellar mass is the same as model D. The simulation is ceased at the age of $50,000~yr$. The density distribution is presented in Figure \ref{fig:modseries4}. In the photoionized region, the temperature of the ionized gas is about $12,000~K$, and the density decreases gradually from $\sim10,000~cm^{-3}$ in the head near the apex to $\sim20~cm^{-3}$ in the tail. In the most part of the H II region, the number density of the ionized gas is lower than $500~cm^{-3}$, but the density is high ($n_e\sim4,000~cm^{-3}$) in the head region. The details of the estimated results in model E are also provided in Table \ref{table_series4}. The estimated density is much higher than the average electron density in the photoionized region. The intensity maps in model E are displayed in Figure \ref{fig:modfimage}. As in the bow shock model, the brightest point in the images is close to the apex of the cometary H II region.  The estimated temperature is also to some extent higher than the average values because of the same reason as in model D.

\begin{figure}[ht!]
\centering
\includegraphics[scale=0.3]{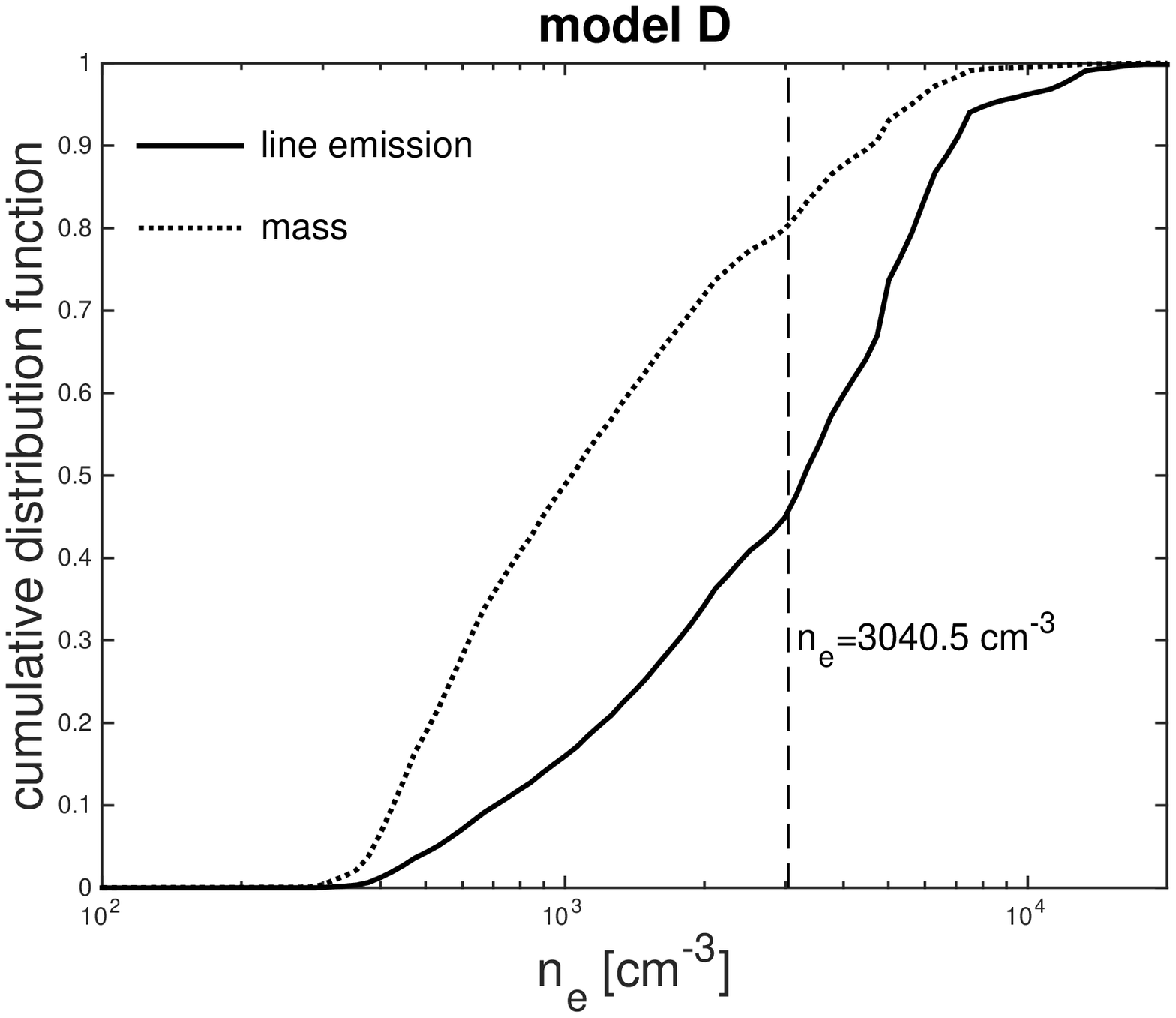}
\includegraphics[scale=0.3]{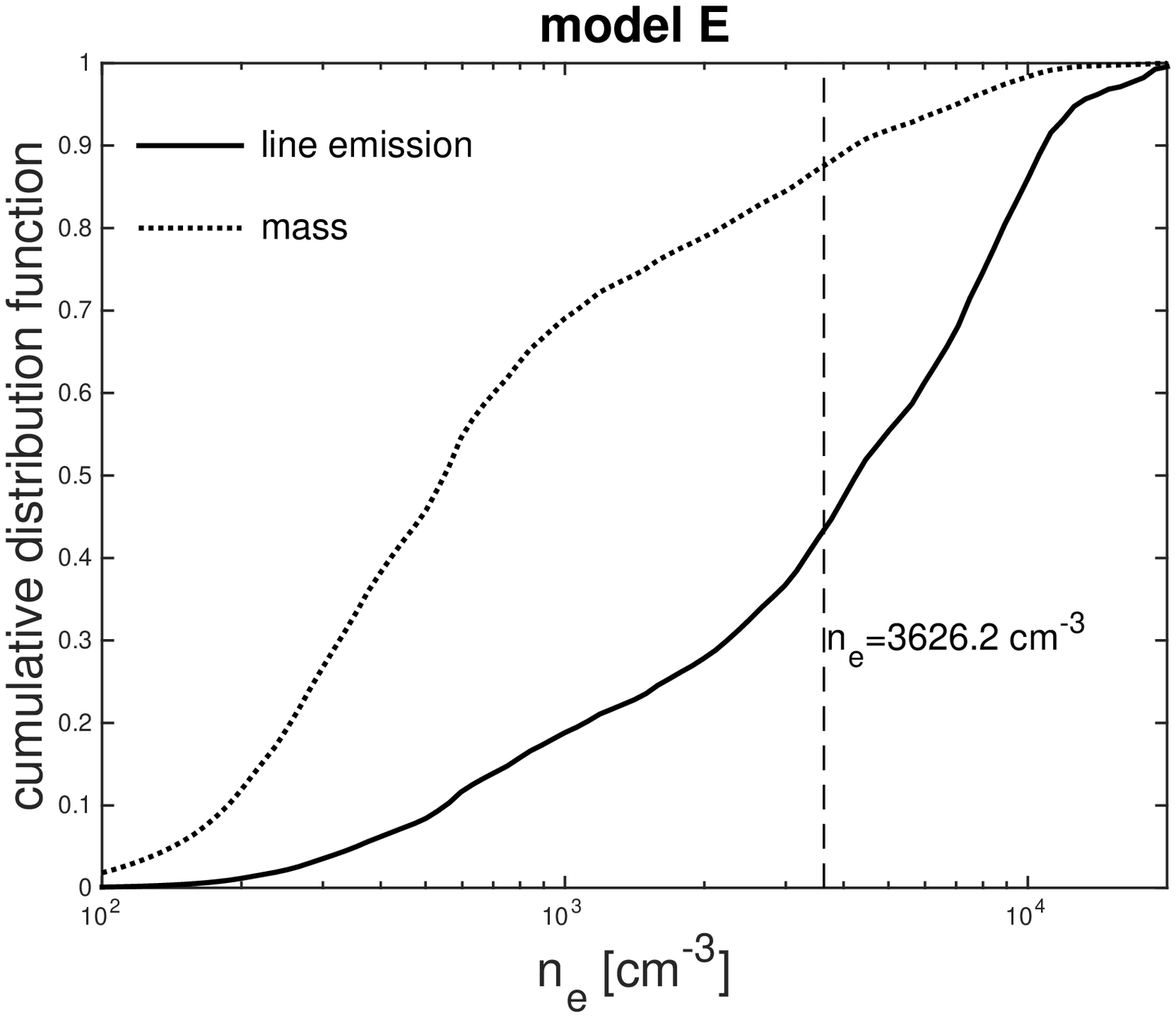}
\includegraphics[scale=0.3]{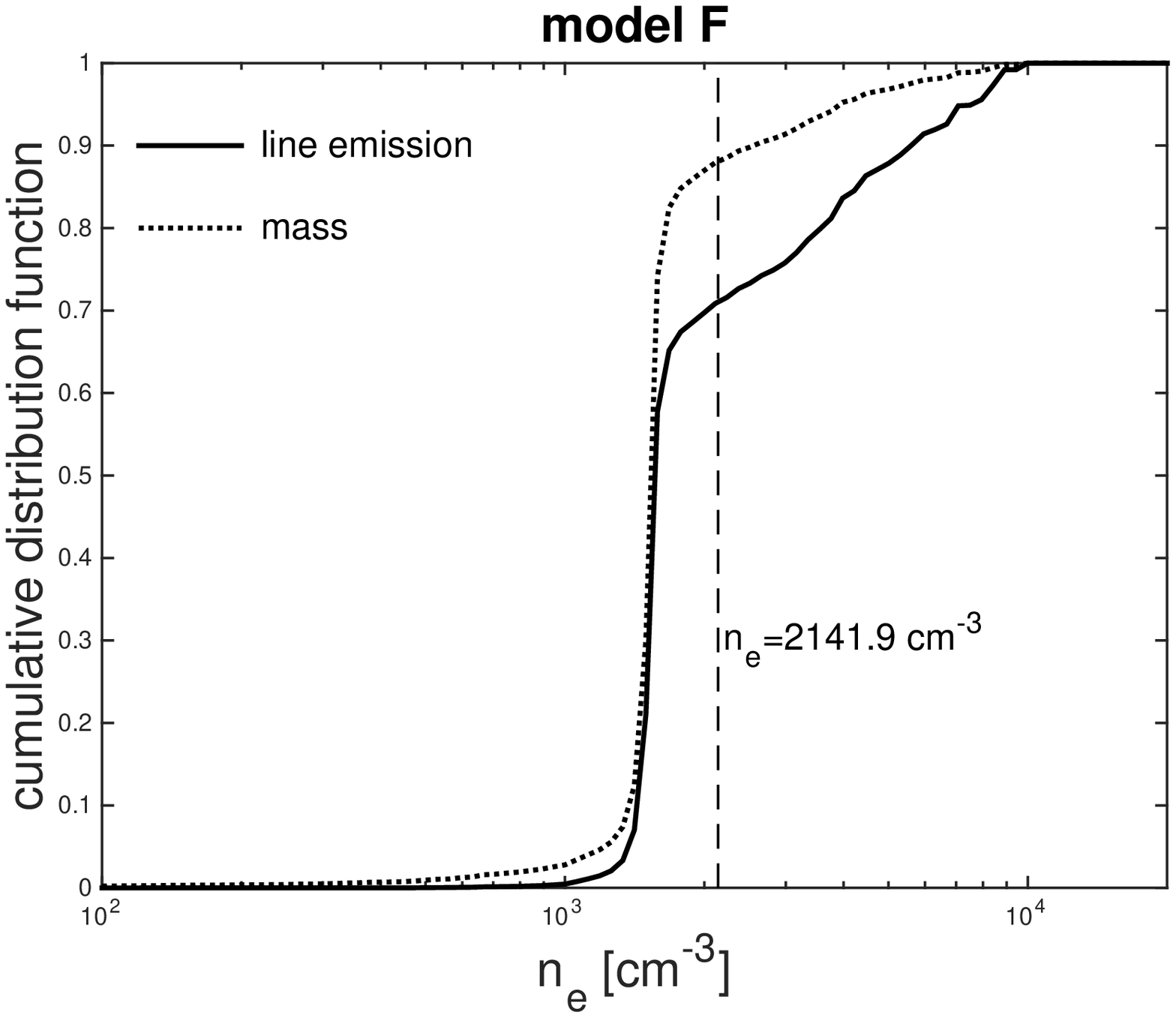}
\caption{The cumulative contribution curves of the hydrogen recombination line emission (H113$\alpha$) and the total mass of the H II region from gas in different electron density intervals for model D (left), E (middle) and F (right). The vertical dashed line indicates the estimated electron number density.}
\label{fig:cdfmodel2}
\end{figure}
\begin{figure}[ht!]
\centering
\includegraphics[scale=0.4]{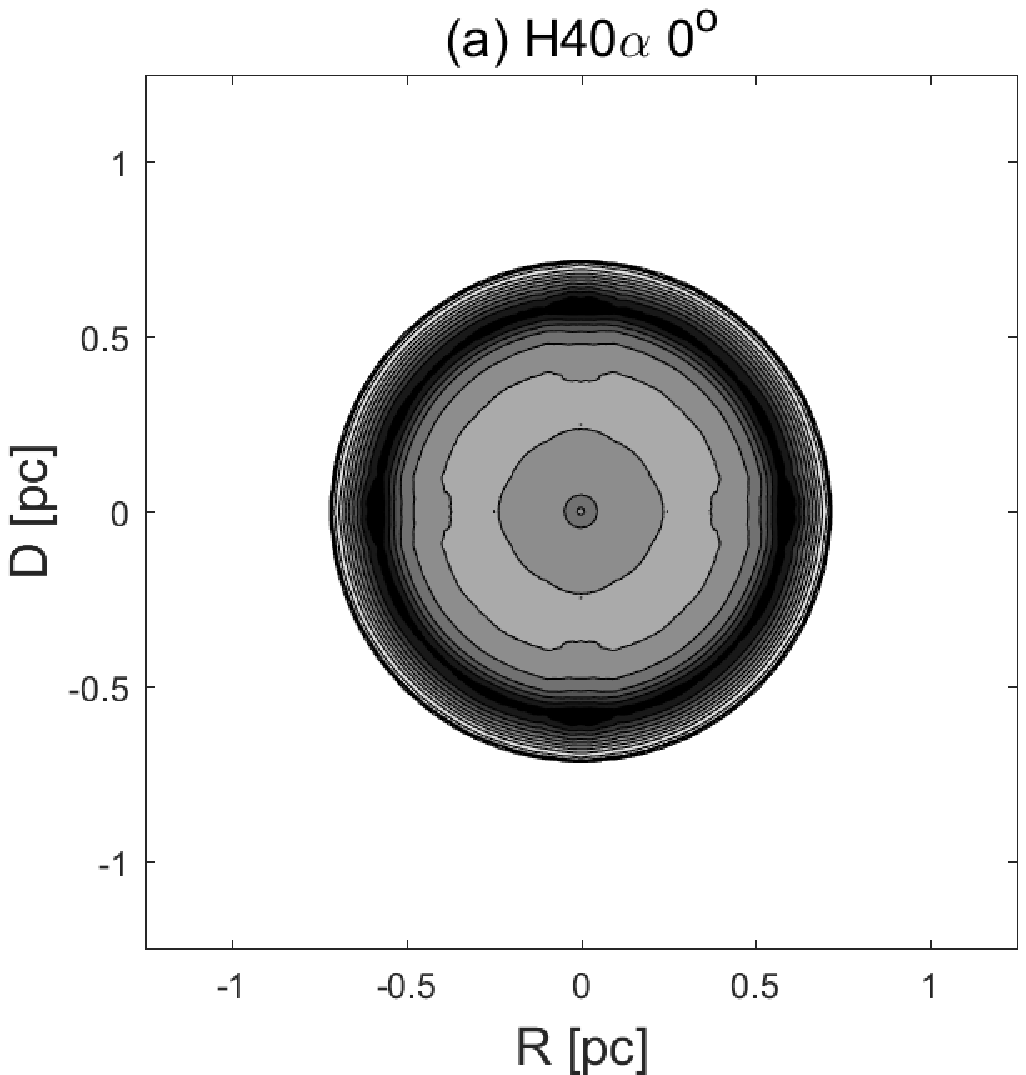}
\includegraphics[scale=0.4]{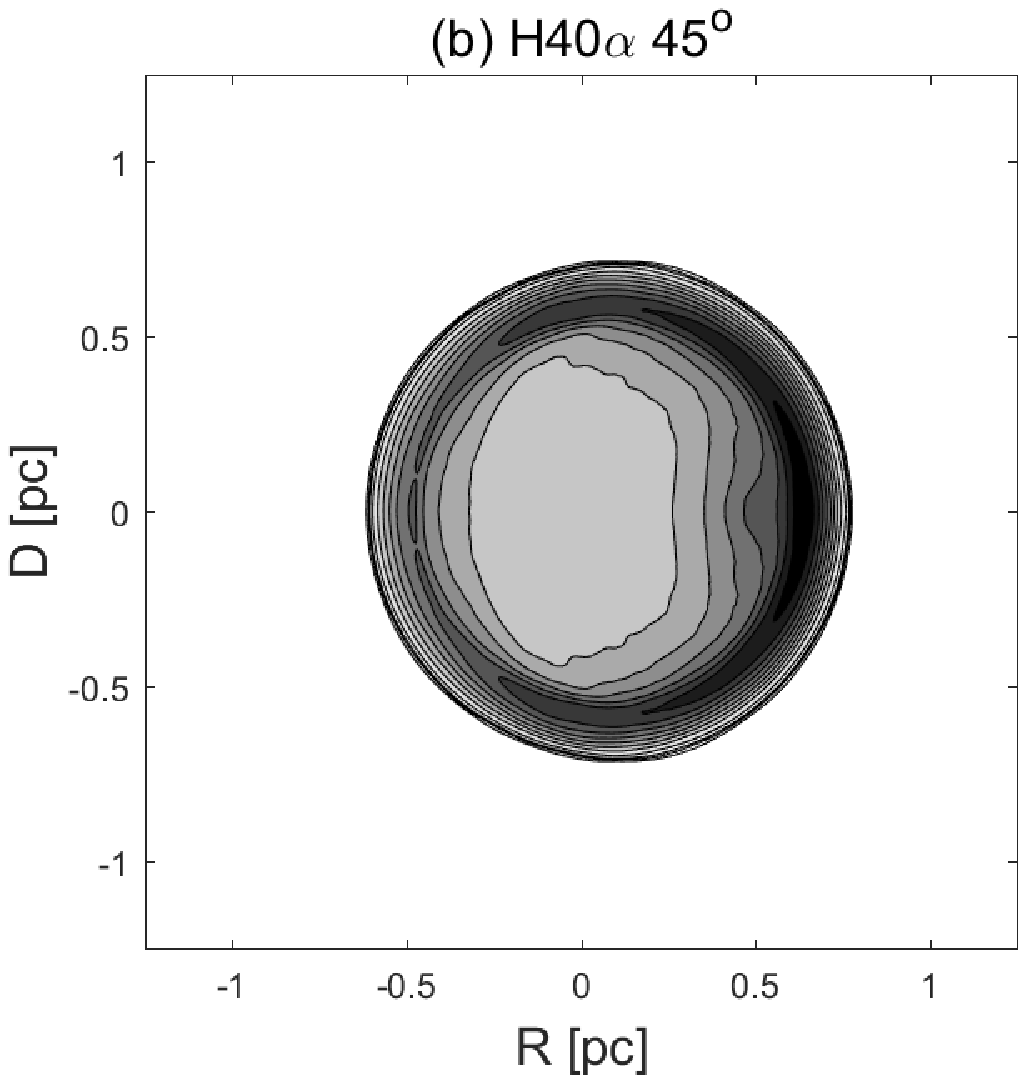}
\includegraphics[scale=0.4]{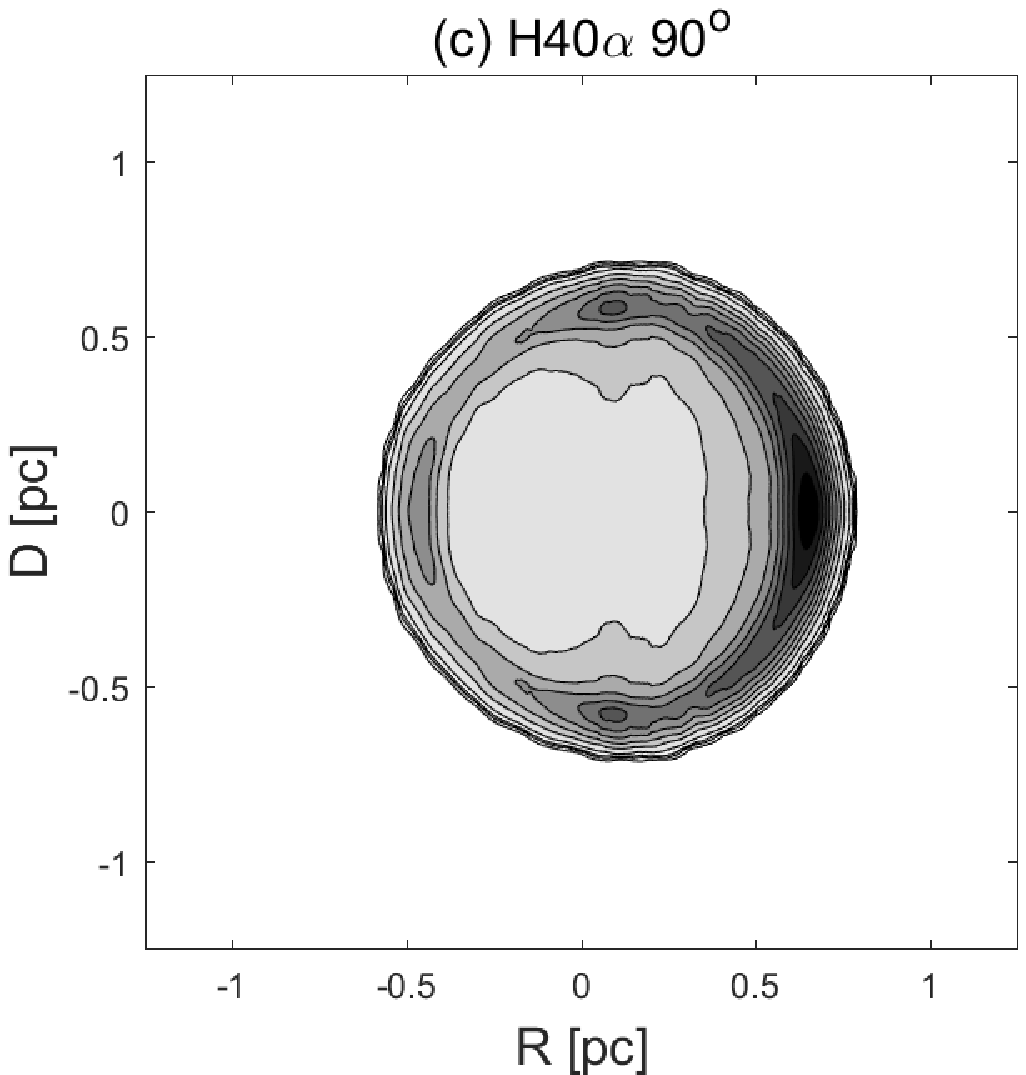}
\includegraphics[scale=0.4]{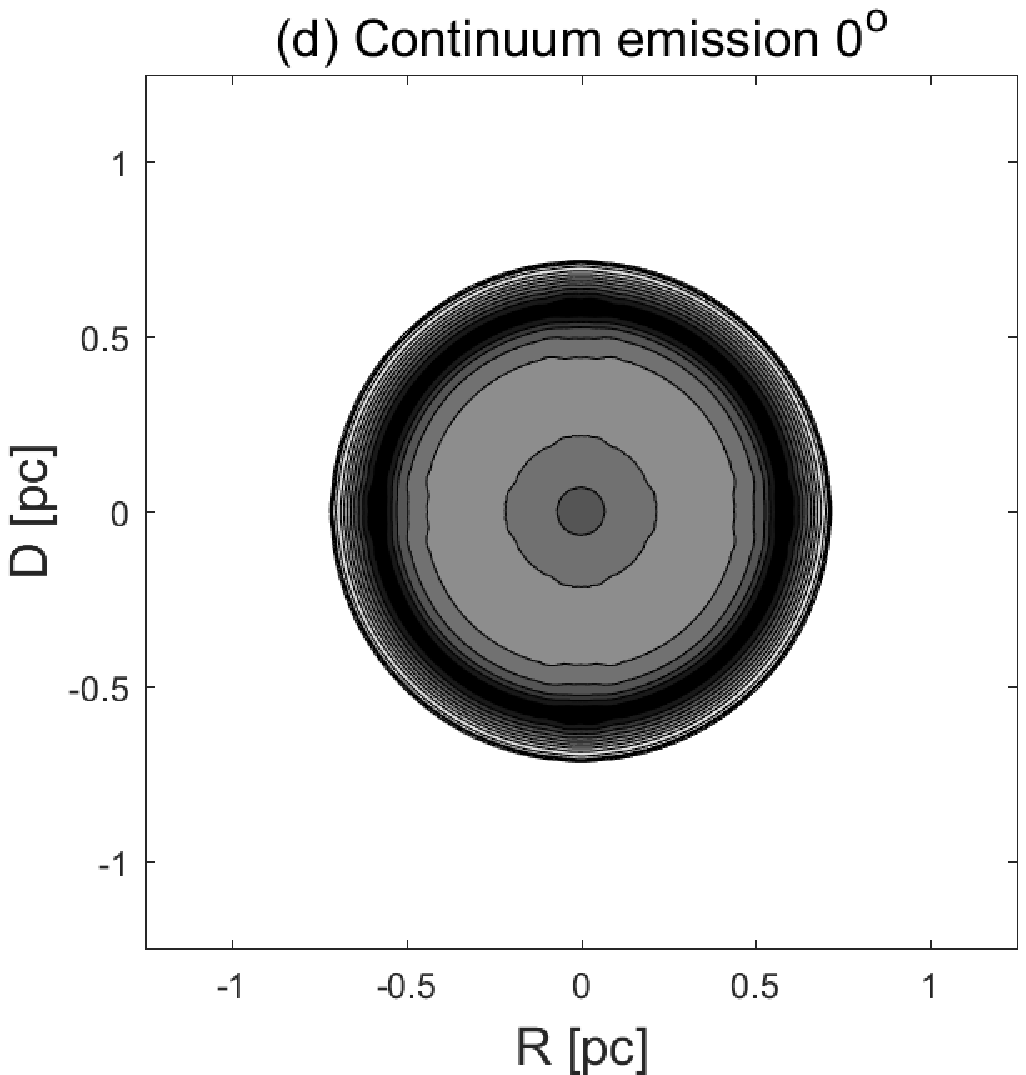}
\includegraphics[scale=0.4]{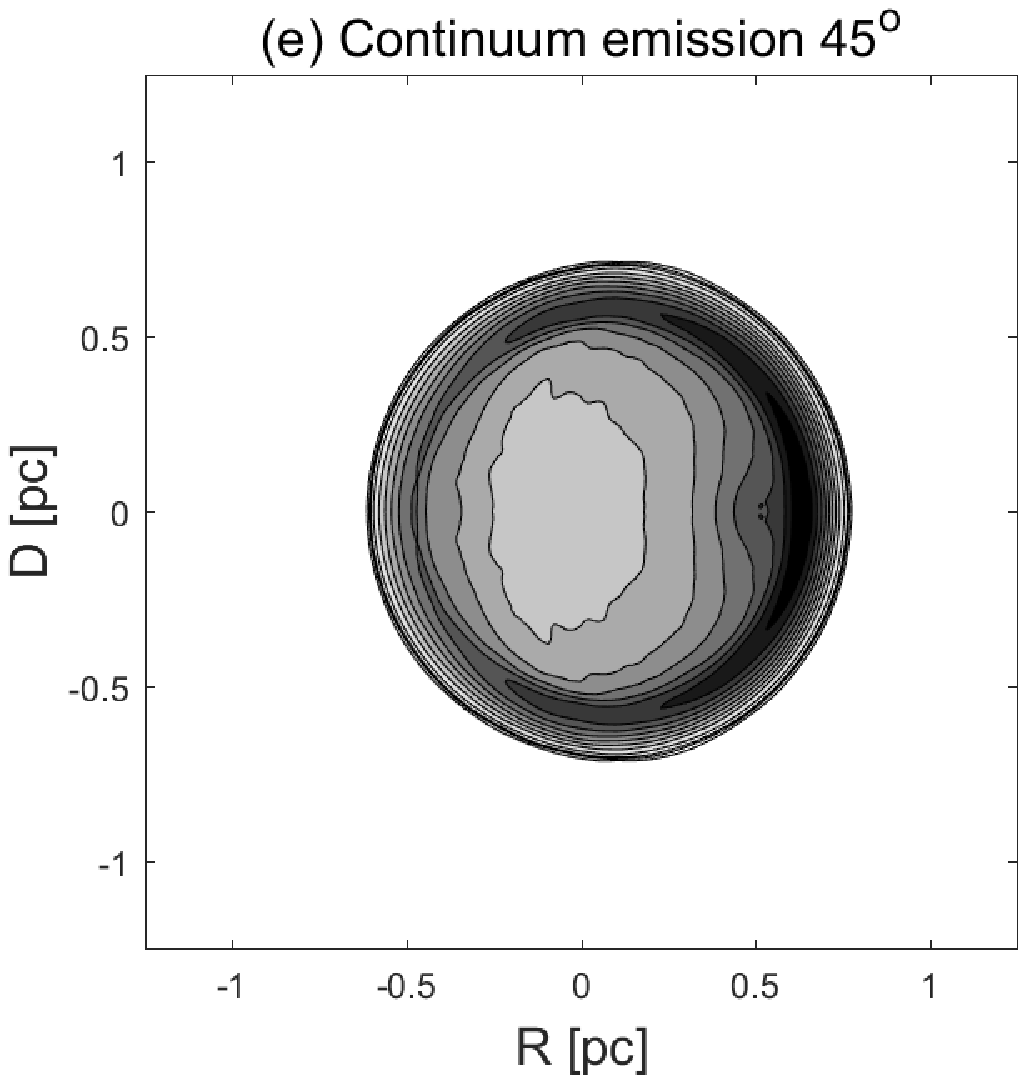}
\includegraphics[scale=0.4]{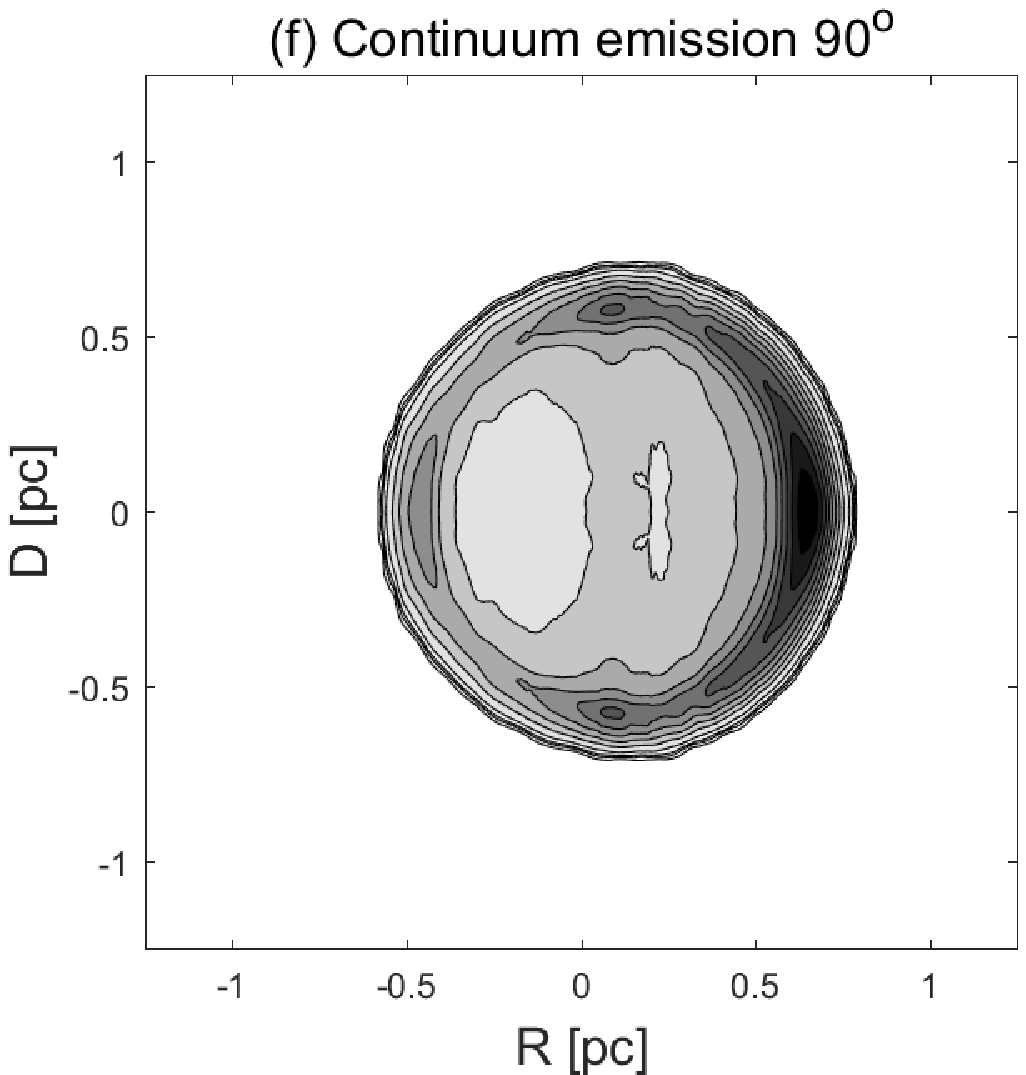}
\caption{The intensity maps of the H40$\alpha$ emission line (a-c) and the continuum emission (d-e) at the frequency of $99.023~GHz$ at three inclination angles ($\theta=0^o,~45^o,~90^o$) in model F. The contour levels are at 1\%, 3\%, 5\%, 10\%, 20\%, 30\%, 40\%, 50\%, 60\%, 70\%, 80\%, 90\% of the emission peaks in each panel.}
\label{fig:moddimage}
\end{figure}

The studies of molecular lines of clouds and radio continuum spectra of compact H II regions show that the density gradients of the molecular clouds where H II regions are born are always approximatively $\rho(r)\propto r^{-\alpha}$. The exponents $\alpha$ are from 1 to 4 \citep{arq85,fra00,tei05}. Here $r$ is the distance to the center. Then we carry out an expansion of the H II region in an pow-law density distribution of $n_H(r)=n_0(r/r_H)^{-2}$ in model F, where $n_0$ is assumed to be $4.0\times10^6~cm^{-3}$, and $r_H=0.05~pc$ is the scale height. The distance from the center of the molecular cloud to the massive star is $1.0~pc$, and the number density at the position of the star is $n=10000~cm^{-3}$. The maximum initial number density is assumed to be $1.0\times10^7~cm^{-3}$ to prevent unreasonable value at the position close to the center. The simulation is ceased at age of $50,000~yr$. The density distributions of all materials and ionized hydrogen in model F are presented in Figure \ref{fig:modseries4}. The $40.9~M_\odot$ star is at the position of $(R,Z)=(0,0)$. Although there is a density gradient in the molecular cloud, the density gradient in the photoionized region is much weakened because of the smoothing effect due to the high sound speed of the ionized gas. This makes the difference between the average electron number density and the flux-weighted average electron number density be much smaller than in model D and E. The intensity maps are shown in Figure \ref{fig:moddimage}. The cumulative distribution curves of the line emission and mass for model D, E and F are presented in Figure \ref{fig:cdfmodel2}. The estimated electron densities in model D and E are both close to the position of cumulative distributions equal to $0.5$, and the cumulative distribution curve in model F is similar to those in spherical H II region models with a stellar wind.


\begin{table}[h]
\centering
\begin{tabular}{|c|c|c|c|c|c|c|}
\hline
\multirow{2}*{Model} & \multirow{2}*{$\bar{T}_e/\bar{T}_e^f$ [K]} & \multirow{2}*{$\bar{n}_e/\bar{n}_e^f$ [$cm^{-3}$]} & \multicolumn{2}{c}{$\sigma/\mu=1\%$} & \multicolumn{2}{|c|}{$\sigma/\mu=3\%$}\\
\cline{4-7}
   &     &      & $\hat{T}$ [K]& $\hat{n}_e$ [$cm^{-3}$] & $\hat{T}$ [K]& $\hat{n}_e$ [$cm^{-3}$] \\
\hline
D & $12189/11992$ & $860.5/3720.1$ & $13902\pm^{1094.8}_{1202.4}$ & $3040.5\pm^{690.2}_{641.7}$ & $12901\pm^{2095.9}_{2858.1}$ & $3286.4\pm^{1499.9}_{1737.6}$ \\
E & $11937/11612$ & $405.1/5299.6$ & $13930\pm^{920.9}_{1105.5}$ & $3626.2\pm^{840.6}_{807.8}$ & $13101\pm^{2494.1}_{2959.6}$ & $3436.2\pm^{1461.6}_{1574.1}$ \\
F & $12245/11877$ & $1570.7/2399.4$ & $12504\pm^{1363.7}_{1321.0}$ & $2141.9\pm^{428.5}_{443.6}$ & $12339\pm^{3256.6}_{3051.5}$ & $2549.2\pm^{1252.7}_{1290.3}$ \\
\hline
\end{tabular}
\caption{The estimated electron temperatures and densities of the model D, E and F are presented. One sigma errors of the estimated values are presented \textbf{under} the assumption of the $1\%$ and $3\%$ uncertainties of $\int S_l d\nu$.}\label{table_series4}
\end{table}

We have also calculated the estimated values in these non-spherical models for different inclination angles $\theta=45^o$ and $90^o$. The estimated values are presented in Table \ref{table_series4ro}, Obviously, the estimated temperatures vary with the inclination angle in Model D and E, this should result from the path-dependence of the Hn$\alpha$ line strengths and the anisotropic structure of cometary H II regions. But the differences between the estimated electron density with different inclination angles are not significant relative to their uncertainties. So we think the electron density estimated by comparing multiple hydrogen radio recombination lines can not be distorted because of the inclination angle for a non-spherical H II region with a significant density gradient, but the estimated electron temperature can.

\begin{table}[h]
\centering
\begin{tabular}{|c|c|c|c|c|c|c|c|c|}
\hline
\multirow{2}*{Model} & \multirow{2}*{$\bar{T}_e/\bar{T}_e^f$ [K]} & \multirow{2}*{$\bar{n}_e/\bar{n}_e^f$ [$cm^{-3}$]} & \multicolumn{2}{c}{$\theta=0^o$} & \multicolumn{2}{|c}{$\theta=45^o$} & \multicolumn{2}{|c|}{$\theta=90^o$} \\
\cline{4-9}
   &     &      & $\hat{T}$ [K]& $\hat{n}_e$ [$cm^{-3}$] & $\hat{T}$ [K]& $\hat{n}_e$ [$cm^{-3}$] & $\hat{T}$ [K]& $\hat{n}_e$ [$cm^{-3}$] \\
\hline
D & $12189/11992$ & $860.5/3720.1$ & $13902\pm^{1094.8}_{1202.4}$ & $3040.5\pm^{690.2}_{641.7}$ & $14492\pm^{1103.4}_{1285.4}$ & $3362.2\pm^{711.6}_{670.7}$ & $13067\pm^{2226.4}_{1662.3}$ & $3313.5\pm^{1051.7}_{969.3}$ \\
E & $11937/11612$ & $405.1/5299.6$ & $13930\pm^{920.9}_{1105.5}$ & $3626.2\pm^{840.6}_{807.8}$ & $14890\pm^{704.9}_{1022.0}$ & $4053.6\pm^{844.2}_{891.4}$ & $12824\pm^{1597.6}_{1749.3}$ & $3615.5\pm^{749.7}_{924.0}$ \\
F & $12245/11877$ & $1570.7/2399.4$ & $12504\pm^{1363.7}_{1321.0}$ & $2141.9\pm^{428.5}_{443.6}$ & $12820\pm^{1321.9}_{1303.7}$ & $2009.1\pm^{389.7}_{387.3}$ & $12856\pm^{1285.8}_{1339.9}$ & $2224.3\pm^{467.3}_{446.0}$ \\
\hline
\end{tabular}
\caption{The estimated electron temperatures and densities with different inclination angles are presented. One sigma errors of the estimated values are presented \textbf{under} the assumption of the $1\%$ uncertainties of $\int S_l d\nu$.}\label{table_series4ro}
\end{table}


\subsection{estimation with measurement of continuum emission}

The results of model A-F display that the the estimated temperature could be deviated significantly from the average temperature if the uncertainty of the frequency-integrated flux is higher than $3\%$. With the uncertainty of that level, the $\sigma/\mu$ of estimated $n_e$ is also high. A method to improve the accuracy is to use an LTE temperature calculated by the line-to-continuum ratio. The ratio of $T_e$ to $T^\ast_e$ is approximately equal to the coefficient $b_m(1-\tau_{\nu,C}\beta/2)$. The continuum optical depth at millimeter wavelengths is very low so that $|\tau_{\nu,C}\beta/2|\ll1$ if it is optically thin at centimeter wavelengths (C band) because the continuum absorption coefficient is approximately $\propto~\nu^{-2.1}$ \citep{alt60}. So we can assume that $T_e/T^\ast_e\approx b_m$ at millimeter wavelengths, and this relation between $T_e$ and $T_e^\ast$ should be satisfied by the electron temperature and density estimated from hydrogen recombination lines. The consideration of this additional condition can significantly improve the precision of the estimation, and can also resolve the deviation of estimated temperature in cometary H II region models. The new estimated properties in Model C3, D and E are presented in Table \ref{table_series5}. They are estimated from 16 Hn$\alpha$ line strengths and the line-to-continuum ratio of H40$\alpha$ to the continuum emission. We assume that the $\sigma$ values of H40$\alpha$ line and the continuum emission are same. Then the $\sigma/\mu$ of continuum emission is lower than that of H40$\alpha$ line since the intensity of the continuum emission is higher.

\begin{table}[h]
\centering
\begin{tabular}{|c|c|c|c|c|c|c|}
\hline
\multirow{2}*{Model} & \multirow{2}*{$\bar{T}_e/\bar{T}_e^f$ [K]} & \multirow{2}*{$\bar{n}_e/\bar{n}_e^f$ [$cm^{-3}$]} & \multicolumn{2}{c}{$\sigma/\mu=3\%$} & \multicolumn{2}{|c|}{$\sigma/\mu=5\%$}\\
\cline{4-7}
   &     &      & $\hat{T}$ [K]& $\hat{n}_e$ [$cm^{-3}$] & $\hat{T}$ [K]& $\hat{n}_e$ [$cm^{-3}$] \\
\hline
C3 & $12289/11915$ & $1759.3/2567.7$ & $11310\pm^{319.9}_{342.9}$ & $2279.3\pm^{811.0}_{866.7}$ & $11358\pm^{618.2}_{603.3}$ & $3071.7\pm^{1395.2}_{1949.7}$ \\
D & $12189/11992$ & $860.5/3720.1$ & $11363\pm^{381.2}_{395.8}$ & $2733.0\pm^{655.4}_{1253.9}$ & $11407\pm^{569.0}_{546.8}$ & $3588.7\pm^{2034.7}_{2466.7}$ \\
E & $11937/11612$ & $405.1/5299.6$ & $11299\pm^{330.3}_{332.5}$ & $3004.6\pm^{1164.1}_{1491.0}$ & $11310\pm^{433.9}_{449.9}$ & $3565.1\pm^{3195.8}_{2390.2}$ \\
\hline
\end{tabular}
\caption{The estimated electron temperatures and densities of the model C3, D and E are presented. One sigma errors of the estimated values are presented \textbf{under} the assumption of the $3\%$ and $5\%$ uncertainties of $\int S_l d\nu$. The line-to-continuum ratio of H40$\alpha$ to the continuum at corresponding frequency is considered in the estimation.}\label{table_series5}
\end{table}

\subsection{the required precision for a accurate estimate}

The results in the H II region models simulated above show that the estimated values are reliable when the uncertainties of the frequency-integrated fluxes of hydrogen recombination lines are lower than $3\%$. And the acceptable uncertainty could be even higher as $5\%$ if one line-to-continuum ratio at millimeter wavelengths is considered. The requirement of the quality of the line observation is a little high. This requirement can be achievable for the many radio telescopes. We use Shanghai Tianma 65-m radio telescope for $4~GHz<\nu<50~GHz$ (C, X, Ku, K and Q bands) and IRAM 30-m telescope for $\nu>80~GHz$ (3-mm band) as an example, and the parameters of these two telescopes are provided in \citet{wan15,wan17} and \citet{kim17}. In the assumption that the distance is $5~kpc$ and the observational time is one hour, the uncertainties of the frequency-integrated luminosities of the hydrogen recombination lines in model C1 are listed in Table \ref{table_uncertianty}. The peak line emission of these lines are also included. There are a large number of sources in the Galaxy with hydrogen recombination lines brighter than those in model C1. So the method described in the current work is practical in estimating the properties of the H II region in the Galaxy.


\begin{table}[h]
\centering
\begin{tabular}{|c|c|c|c|}
\hline
Line & Luminosities [$erg~s^{-1}$] & Peak line emission [mJy] & error [\%] \\
\hline
H40$\alpha$ & $2.98\times10^{29}$ & 1116.9 & 0.95 \\
H52$\alpha$ & $6.32\times10^{28}$ & 515.6 & 0.42 \\
H63$\alpha$ & $2.11\times10^{28}$ & 308.6 & 0.69 \\
H71$\alpha$ & $1.10\times10^{28}$ & 226.1 & 0.60 \\
H80$\alpha$ & $5.62\times10^{27}$ & 165.2 & 0.98 \\
H90$\alpha$ & $2.89\times10^{27}$ & 120.4 & 1.33 \\
H100$\alpha$ & $1.59\times10^{27}$ & 90.4 & 1.28 \\
H113$\alpha$ & $0.80\times10^{27}$ & 64.6 & 2.13 \\
\hline
\end{tabular}
\caption{The properties and the uncertainties of hydrogen recombination lines in model C1. A distance of $5~kpc$ and an 1-hour observation time are assumed. The errors of the luminosities of the H40$\alpha$ line are calculated by the instrumental parameters of IRAM 30-m telescope. The errors of the other hydrogen recombination lines are calculated by the parameters of Shanghai Tianma 65-m radio telescope.}\label{table_uncertianty}
\end{table}

\section{Conclusions} \label{sec:conclusion}

In the paper, we introduced a method about estimating the electron temperature and number density of an H II region by using hydrogen RRLs. We derive line flux ratios at low optical depth conditions. We use evolutionary hydrodynamical models of H II regions to test accuracy of the method under a variety of conditions. The results show that the method can estimate the properties of the H II regions in the Galaxy to acceptable accuracy. The advantage of this method is that the density and temperature measurements can be carried out with hydrogen recombination line observations on single dish telescopes.  If the hydrogen RRLs observations are accurate enough ($\sigma/\mu <3\%$), the estimation can be treated without continuum observations. The disadvantage of this method is the requirement of high-quality line observational data ($\sigma/\mu <5\%$), but this is achievable for many radio telescopes. Our conclusions are summarized as follows.


1. If there is a gas density gradient in the H II region, the estimated electron temperature and density are more representative of the properties of the ionized gas in the relatively high-density region. Such as in a cometary H II region, the estimated electron density could be much higher than the average electron density and represent the density in the head region due to its large density gradient. For the H II region with a low-density stellar wind bubble, the estimated properties show the condition in the photoionized region because the line emission from the stellar bubble can be negligible.


2. By using only multiple hydrogen RRLs to estimate properties of H II regions, high-quality observational data are necessary. Based on the results in this paper, the average uncertainty of the frequency-integrated fluxes $\int S_l d\nu$ of different hydrogen recombination lines should be lower than $3\%$ for accurate estimated values.

3. It is necessary to observe the hydrogen recombination lines from at least 3 frequency bands in order to obtain an accurate estimate.

4. For a cometary H II region with a high density gradient, the estimated electron temperature may be several thousands Kelvins higher or less than the real value. But a reasonable estimated value of the electron density can be achieved.

5. The accuracies of the estimated temperature and density can be significantly increased if a line-to-continuum ratio at millimeter wavelengths is used in the estimation. The electron temperature in a cometary H II region can also be correctly approached by this method.



\acknowledgements
The work is partially supported by Natural Science Foundation of China No.11590782, No.11421303 and No.11473007. F.-Y. Zhu thanks researchers at Shanghai Astronomical Observatory and Purple Mountain Observatory for kindly answering questions about observations with radio telescopes.

\end{document}